\documentclass[aps,prc,twocolumn,preprintnumbers,
               nofootinbib,float,superscriptaddress,longbibliography]{revtex4-1}

\usepackage[latin9]{inputenc}
\usepackage{hyperref}
\hypersetup{
  colorlinks=true,        
  linkcolor=blue,         
  citecolor=cyan,         
}
\usepackage{breakurl}
\usepackage{graphicx}
\usepackage{epsf}
\usepackage{epsfig}
\usepackage{amssymb,amsmath}
\usepackage[usenames]{color}
\usepackage{amssymb}
\usepackage{times}
\usepackage{comment}

\usepackage[normalem]{ulem}

\newcommand\ru{{^{96}_{44}\textrm{Ru}}}
\newcommand\zr{{^{96}_{40}\textrm{Zr}}}

\newcommand\snn{\sqrt{s_\text{NN}}}



\begin{document}

\preprint{This line only printed with preprint option}

\title{Hydrodynamic simulations of directed flow for light hadrons in Au+Au and isobar collisions at $\snn=$ 200 GeV}

\author{Jing Jing}
\affiliation{Department of Physics and Electronic-Information Engineering, Hubei Engineering University, Xiaogan, Hubei, 432000, China}

\author{Ze-Fang Jiang}
\email{jiangzf@mails.ccnu.edu.cn}
\affiliation{Department of Physics and Electronic-Information Engineering, Hubei Engineering University, Xiaogan, Hubei, 432000, China}
\affiliation{Institute of Particle Physics and Key Laboratory of Quark and Lepton Physics (MOE), Central China Normal University, Wuhan, Hubei, 430079, China}

\author{C. B. Yang}
\affiliation{Institute of Particle Physics and Key Laboratory of Quark and Lepton Physics (MOE), Central China Normal University, Wuhan, Hubei, 430079, China}

\author{Ben-Wei Zhang}
\affiliation{Institute of Particle Physics and Key Laboratory of Quark and Lepton Physics (MOE), Central China Normal University, Wuhan, Hubei, 430079, China}

\begin{abstract}
Using a (3+1)-D hydrodynamic model CLVisc, we study the directed flow ($v_{1}$) of light hadrons produced in Au+Au, Ru+Ru and Zr+Zr collisions at $\snn=$ 200 GeV.
The evolution of tilted energy density, pressure gradient and radial flow along the $x$-direction are systematically investigated.
Counter-clockwise tilt of initial fireball is shown to be a vital source of the directed flow for final light hadrons.
A good description of directed flow is provided for light hadrons in central and mid-central Au+Au and isobar collisions at RHIC.
Our numerical results show a clear system size dependence for light hadron $v_{1}$ across different collision systems.
We further study the effect of nuclear structure on the directed flow and find that the $v_{1}$ for light hadrons is insensitive to the nuclei with quadrupole deformation.
\end{abstract}
\maketitle
\date{\today}

\section{Introduction}
\label{emsection1}

High energy nucleus-nucleus collisions preformed at the Relativistic Heavy-Ion Collider (RHIC) and the Large Hadron Collider (LHC) suggest that a novel color-deconfined QCD matter (Quark-Gluon Plasma, known as QGP), is created in the reaction region.
The azimuthal asymmetric flows (collective flow) of the observed hadrons in various collision systems~\cite{PHENIX:2003qra,ALICE:2010suc,CMS:2012zex} are important phenomena in the study of the strongly interacting nature of the QGP,
such as directed flow $v_{1}$, elliptic flow $v_{2}$, triangular flow $v_3$, etc. Collective flows have been successfully described by relativistic hydrodynamic models~\cite{Ollitrault:1992bk,Rischke:1995ir,Sorge:1996pc,Bass:1998vz,Aguiar:2001ac,Shuryak:2003xe,Gyulassy:2004zy,Broniowski:2007ft,
Andrade:2008xh,Hirano:2009ah,Schenke:2010rr,Qiu:2011iv,Heinz:2013th,Huovinen:2013wma,Gale:2013da,Bozek:2013uha,Qin:2013bha,Dusling:2015gta,Romatschke:2017ejr,
Weller:2017tsr,Zhao:2020wcd}, and the shear viscosity ratio ($\eta_{v}/s$) extracted from the experimental data seems to be small~\cite{Song:2010mg,Bernhard:2019bmu}.

The directed flow ($v_{1}$) is one of earliest observables for investigating the collectivity properties in heavy ion collisions~\cite{Gyulassy:1981nq,Gustafsson:1984ka,Lisa:2000ip}. It is defined by the first-order Fourier coefficient of the final-hadron azimuthal distribution and has been widely investigated at both RHIC and LHC~\cite{Voloshin:1994mz,Bilandzic:2010jr,STAR:2004jwm,STAR:2014clz,STAR:2017okv,STAR:2019clv,ALICE:2019sgg,STAR:2019vcp}.
Many studies suggested that the directed flow is generated at very early stage in the nuclear collisions, whose typical time scale is nearly $2R/\gamma$, where $R$ and $\gamma$ are the nuclear radius and Lorentz boost factor, respectively~\cite{Gyulassy:1981nq,Sorge:1996pc,Singha:2016mna}.
Such a time scale is even shorter than the producing time of the elliptic flow $v_{2}$. Therefore, directed flow $v_1$ could be a useful probe to investigate the medium distribution and nucleon flow at the initial stage in nucleus-nucleus collisions~\cite{Ollitrault:1992bk,Voloshin:1994mz,Nara:2016phs,Chatterjee:2017ahy,Singha:2016mna,Zhang:2018wlk,Guo:2017mkf,Parida:2022lmt}.
Various mechanisms could contribute to the directed flow of light hadron. Model calculations suggest that the $v_1$ could depend on the deformation of the initial fireball geometry, the initial baryon density distribution, the flow velocity field, the equation of state of QGP medium, the external electromagnetic fields and also the final hadronic rescatterings~\cite{Adil:2005qn,Bozek:2010bi,Chen:2019qzx,Shen:2020jwv,Ryu:2021lnx,Chatterjee:2017ahy,Chatterjee:2018lsx,Beraudo:2021ont,Bozek:2022svy}, although their exact quantitative contributions are still open and unsolved questions.

The recent isobar run of collisions of both $\ru$ and $\zr$ \footnote{We refer to $\ru$ and $\zr$ as Ru and Zr in the subsequent sections.} at $\snn=200$ GeV by the STAR Collaboration at RHIC~\cite{STAR:2021mii} has a special motivation to search for the chiral magnetic effect (CME)~\cite{Fukushima:2008xe,Skokov:2009qp,Deng:2016knn,Zhao:2019crj}.
However, the isobar blind analysis did not provide predefined CME signal so far, but surprisingly found that Ru+Ru collisions provided a higher particle yields than Zr+Zr collisions as well as larger elliptic flow ($v_{2}$), albeit having a smaller triangular flow ($v_{3}$)~\cite{Nijs:2021kvn,Zhang:2022fou}. Those differences of experimental results at STAR indicate a difference in nuclear structure (geometry shape) whose observable effects are seemingly larger than those induced by the difference of the electric charge between Ru and Zr~\cite{STAR:2021mii}, and have been investigated from nuclear structure analyses by many studies~\cite{Nijs:2021kvn,Zhang:2021kxj,Jia:2021oyt,Xu:2021uar,Li:2022bhl,Zhao:2022grq,Ma:2022dbh,Jia:2022qrq}. Therefore, it is of great interest to analyse a detailed comparison between Ru and Zr within a uniform QGP evolution framework, and identify the main features of the nuclear structure (or system size) that contribute to the directed flow $v_{1}$ of identified final-state light hadrons.

In this work, we utilize a (3+1)-D viscous hydrodynamic model (CLVisc) with a tilted initial condition~\cite{Pang:2016igs,Pang:2018zzo,Wu:2018cpc} to investigate the origin of directed flow of light hadrons produced in Au+Au and isobar collisions at $\snn=200$ GeV, with nuclear structure parameters for Ru and Zr from the energy density functional theory (DFT) calculations~\cite{Xu:2017zcn,Li:2018oec,Xu:2021vpn}. The correlation between the fireball structure of the initial state and the directed flow coefficient of the final state light hadrons is presented. Our numerical results find that the tilted initial energy density profile for different nuclei (Au, Ru and Zr) yields different nonzero average pressure gradients $-\langle \partial_x P\rangle$ at forward/backward space-time rapidity, which further induces a negative slope of the average flow velocity $\langle v_x \rangle$ with respect to the space-time rapidity $\eta_s$, and finally the same size and sign of $v_1$ {\it vs.} $\eta$, which is consistent with the experimental data in Au+Au and isobar collisions at $\snn=200$ GeV~\cite{Abelev:2008jga,STAR:2021mii,qm2022ruzr}. Our calculation shows a clear system size dependence for light hadron $v_{1}$ across different collision systems.
In the end, we compare the directed flow ($v_{1}$) between three different nuclear structures with quadrupole deformation ($\beta_{2}$) for Ru and Zr nuclei.

This article is organized as follows. In Sec.~\ref{v1section2}, we present the rapidity-dependent energy density distribution in Au+Au and isobar collisions and their impacts on the pressure gradient and flow velocity with respect to time during the hydrodynamic simulations.
In Sec.~\ref{v1section3}, we present the directed flow of light hadrons from our hydrodynamic calculation and study its dependence on the nuclear structure. We present a brief summary in Sec.~\ref{v1section4}.

\section{The model framework}
\label{v1section2}
\subsection{Parametrization of longitudinal profiles for energy density}
\label{v1subsect2}

In order to investigate the directed flow of light hadrons in Au+Au and isobar collisions, following our previous works~\cite{Jiang:2021foj,Jiang:2021ajc,Jiang:2022uoe,Li:2022pyw},
we start with the initial energy density distribution for nuclei Au, Ru and Zr.
Their impacts on the pressure gradient and flow velocity with respect to time will be investigated using the (3+1)-D hydrodynamic model CLVisc.

Based on the Woods-Saxon (WS) distribution, the nucleus thickness function is defined as
\begin{equation}
\begin{aligned}
T(x,y)=\int_{-\infty}^{\infty}dz\frac{n_{0}}{1+\exp\left[\frac{r-R_{0}(1+\beta_{2}Y^{0}_{2}(\theta))}{d}\right]},
\label{eq:thicknessf}
\end{aligned}
\end{equation}
where $r=\sqrt{x^{2}+y^{2}+z^{2}}$ is the radial position, $x,~y,~z$ are the space coordinates, $\theta$ is the polar angle in their rest frame, $d$ is the surface diffusiveness parameter,
$\beta_{2}$ is the quadruple deformity of nucleus, $Y_{2}^{0}(\theta)=\frac{1}{4}\sqrt{\frac{5}{\pi}}(3\cos^{2}\theta-1)$, and $R_{0}$ is the radius of the nuclear, which depends on the nucleus species.
The values of the parameters used for Au, Ru and Zr in the current study are listed in Tab.~\ref{t:parameters}.
At present, nuclear density distribution of Ru and Zr are not accurately confirmed, because there are numbers of setups of parameters ($R_{0}$, $d$ and $\beta_{2}$) from different experiments and models.
In this work, following the pioneering work~\cite{Xu:2017zcn,Li:2018oec,Xu:2021vpn} and the STAR experiment~\cite{STAR:2021mii}, we first adopt the nuclei sets from recent calculations based on energy density functional theory (DFT)~\cite{Xu:2017zcn,Li:2018oec,Xu:2021vpn,Deng:2018dut} to study the light hadron directed flow.
The other two sets will be discussed later when we investigate the effect of nuclear structure on the directed flow $v_1$ of final light hadrons.
\begin{table}[!h]
\begin{center}
\caption{\label{t:parameters} Nuclear parameters used in the Woods-Saxon distribution for Au, Ru and Zr~\cite{Loizides:2017ack,STAR:2021mii}.}
\begin{tabular}{ l | c | c | c | c }
\hline\hline
Nucleus            & $n_{0}$ [1/fm$^{3}$]    & $R_{0}$~[fm]    & $d$ [fm]  & $\beta_{2}$   \\ \hline
$^{197}_{79}$Au    & 0.17                    & 6.38        & 0.535     & 0.0 \\
$^{96}_{44}$Ru     & 0.17                    & 5.067       & 0.500      & 0.0 \\
$^{96}_{40}$Zr     & 0.17                    & 4.965        & 0.556      & 0.0 \\
\hline\hline
\end{tabular}
\end{center}
\end{table}

Considering the projectile and target nuclei propagating along $\pm \hat{z}$ direction with the impact parameter $\mathbf{b}$, the corresponding thickness function can be written as
\begin{equation}
\begin{aligned}
T_{+}(\mathbf{x}_\text{T})=T(\mathbf{x}_\text{T}-\mathbf{b}/2),~~~~T_{-}(\mathbf{x}_\text{T})=T(\mathbf{x}_\text{T}+\mathbf{b}/2),
\label{eq:t+}
\end{aligned}
\end{equation}
where $\mathbf{x}_\text{T}=(x,y)$ is the transverse plane coordinate.
The density distributions of the participant nucleons are then given by
\begin{equation}
\begin{aligned}
T_{1}(\mathbf{x}_\text{T})=T_{+}(\mathbf{x}_\text{T})\left\{1-\left[1-\frac{\sigma_\text{NN} T_{-}(\mathbf{x}_\text{T})}{A}\right]^{A}\right\},
\label{eq:t1}
\end{aligned}
\end{equation}
\begin{equation}
\begin{aligned}
T_{2}(\mathbf{x}_\text{T})=T_{-}(\mathbf{x}_\text{T})\left\{1-\left[1-\frac{\sigma_\text{NN} T_{+}(\mathbf{x}_\text{T})}{A}\right]^{A}\right\},
\label{eq:t2}
\end{aligned}
\end{equation}
where $A$ is the nuclei mass number, $\sigma_\text{NN}$ = 42 mb is the inelastic nucleon-nucleon scattering cross section at $\snn=200$ GeV~\cite{Loizides:2017ack}.
The centrality bin in different nuclear collisions are determined by the impact parameter $\mathbf{b}$~\cite{Loizides:2017ack}.

Since the right/left-moving wounded nucleons (as shown in Fig.~\ref{f:auau200ed}) are expected to emit more particles at forward/backward rapidity,
we assume this can be constructed by introducing a deformation mechanism into the weight function $W_\text{N}$~\cite{Jiang:2021foj},
in which a tilted fireball is introduced to describe the observed charged particle directed flow at RHIC and LHC energy region~\cite{Bozek:2011ua,Jiang:2021foj}.
In our earlier studies~\cite{Jiang:2021foj,Jiang:2021ajc,Jiang:2022uoe}, a monotonic function ($T_{1}(x,y)+T_{2}(x,y)$) was modified to obtain the asymmetry between the forward and backward nuclei as follow,
\begin{equation}
\begin{aligned}
W_\text{N}(x,y,\eta_{s})&=T_{1}(x,y)+T_{2}(x,y) \\
+&H_{t}[T_{1}(x,y)-T_{2}(x,y)]\tan\left(\frac{\eta_{s}}{\eta_{t}}\right),
\label{eq:mnccnu}
\end{aligned}
\end{equation}
where $H_{t}$ reflects the strength of imbalance at the forward and backward rapidities,
the function $\tan (\eta_{s}/\eta_{t})$ produces the deformation of the initial energy density distribution along the longitudinal direction, and $\eta_{t}=8.0$ is utilized for all the collision systems~\cite{Jiang:2021foj}.
In different centrality classes, the baryon stopping effect is different,
hence the contribution from the forward and backward nuclei are imbalanced.
In our model, parameter $H_{t}$ reflects the strength of baryon stopping effect and it depends on the centrality.
The value of $H_{t}$ is extracted from the experimental data.

The total weight function $W(x,y,\eta_{s})$ is defined as
\begin{equation}
\begin{aligned}
W(x,y,\eta_{s})=\frac{0.95 W_\text{N}(x,y,\eta_{s})+0.05 n_\text{BC}(x,y)}{\left[0.95 W_\text{N}(0,0,0)+0.05 n_\text{BC}(0,0)\right]|_{\mathbf{b}=0}}.
\label{eq:wneta}
\end{aligned}
\end{equation}
Here, the number of binary collisions $n_\text{BC}(x,y)$ is defined as~\cite{Pang:2018zzo,Loizides:2017ack}
\begin{equation}
\begin{aligned}
n_\text{BC}(x,y)=\sigma_\text{NN}T_{+}(x,y)T_{-}(x,y).
\label{eq:nbc}
\end{aligned}
\end{equation}

The initial energy density $\varepsilon(x,y,\eta_{s})$ is given by~\cite{Pang:2018zzo}
\begin{equation}
\begin{aligned}
\varepsilon(x,y,\eta_{s})=K \cdot W(x,y,\eta_{s}) \cdot H(\eta_{s}),
\label{eq:ekw}
\end{aligned}
\end{equation}
where $K$ is a normalization factor and determined by the multiplicity density distribution ($dN_{\textrm{ch}}/d\eta$) of soft particles. A function
\begin{equation}
\begin{aligned}
H(\eta_{s})=\exp\left[-\frac{(|\eta_{s}|-\eta_{w})^{2}}{2\sigma^{2}_{\eta}}\theta(|\eta_{s}|-\eta_{w}) \right]
\label{eq:heta}
\end{aligned}
\end{equation}
is introduced to describe the plateau structure of the rapidity distribution of emitted hadrons at mid-rapidity,
in which $\eta_{w}=1.3$ determines the width of the central rapidity plateau
while $\sigma_{\eta}=1.5$ determines the width (spread) of the Gaussian decay from the plateau region~\cite{Pang:2018zzo}.

In Tab.~\ref{t:modelparameters}, we summarize the parameters of initial conditions that are tuned to provide reasonable descriptions of the charged particle yields in the most central collisions~\cite{Pang:2018zzo}, as will be shown in Fig.~\ref{f:auau200dndeta}. They include the overall normalization factor ($K$), the initial time of the hydrodynamic evolution ($\tau_{0}$), the impact parameters ($b$) and the tilted parameters ($H_{t}$).

\begin{table}[!h]
\begin{center}
\caption{\label{t:modelparameters} Parameters used in hydrodynamic simulations between different nuclei~\cite{Pang:2018zzo,Loizides:2017ack,Jiang:2021foj,Jiang:2021ajc}.}
\begin{tabular}{c | c |c| c }
\hline
\hline
  Parameters                      & Au+Au             & Ru+Ru   & Zr+Zr  \\
\hline
 $K$ (GeV/fm$^{3}$ )                   & 35.5            &23.0   &23.0   \\
 $\tau_{0}$ (fm/c)                & 0.6           &0.6   &0.6\\
$b$ (fm)                & 2.4           &2.1   &2.1\\
$H_{t}$                 & 1.0           &1.0   &1.0\\
\hline
\hline
\end{tabular}
\end{center}
\end{table}

In present work, we set the initial fluid velocity in the transverse and space-time rapidity directions following the Bjorken approximation, i.e., $v_{x} = v_{y} =0$ and $v_{z} = z/t$.

\subsection{Tilted energy density, eccentricity and pressure gradient}

\begin{figure}[!tbp]
\begin{center}
\includegraphics[width=0.75\linewidth]{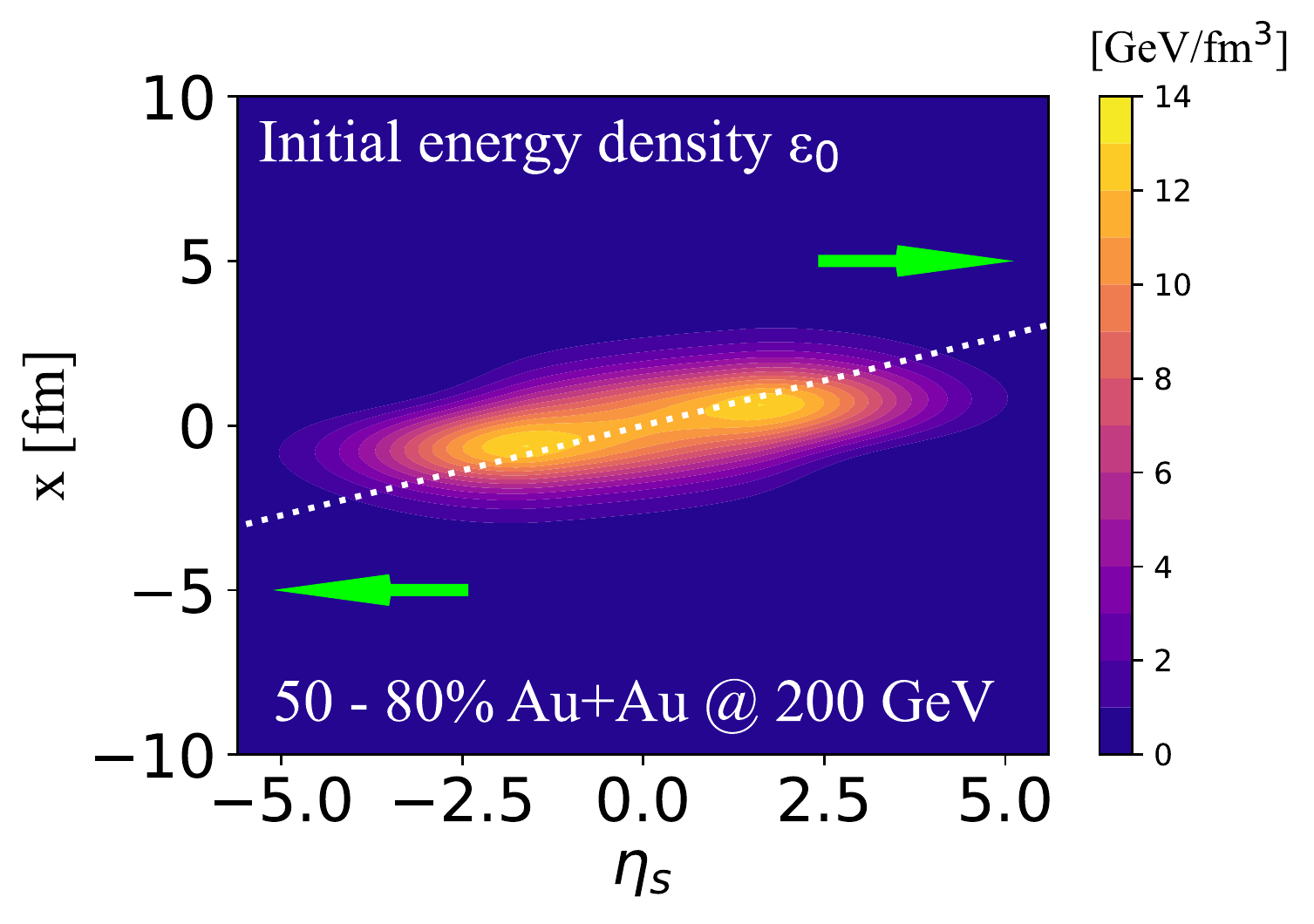}~\\
\includegraphics[width=0.75\linewidth]{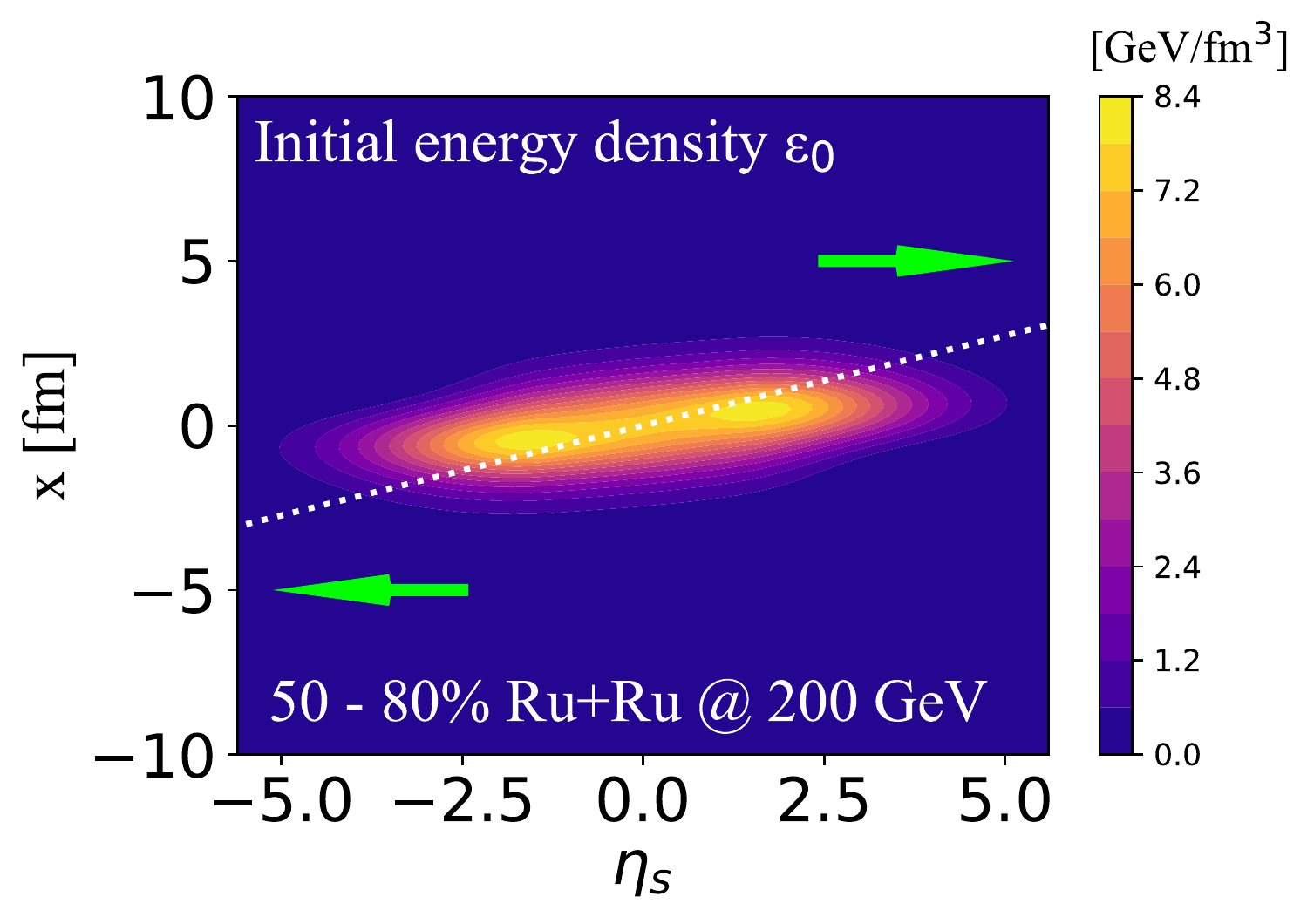}~\\
\includegraphics[width=0.75\linewidth]{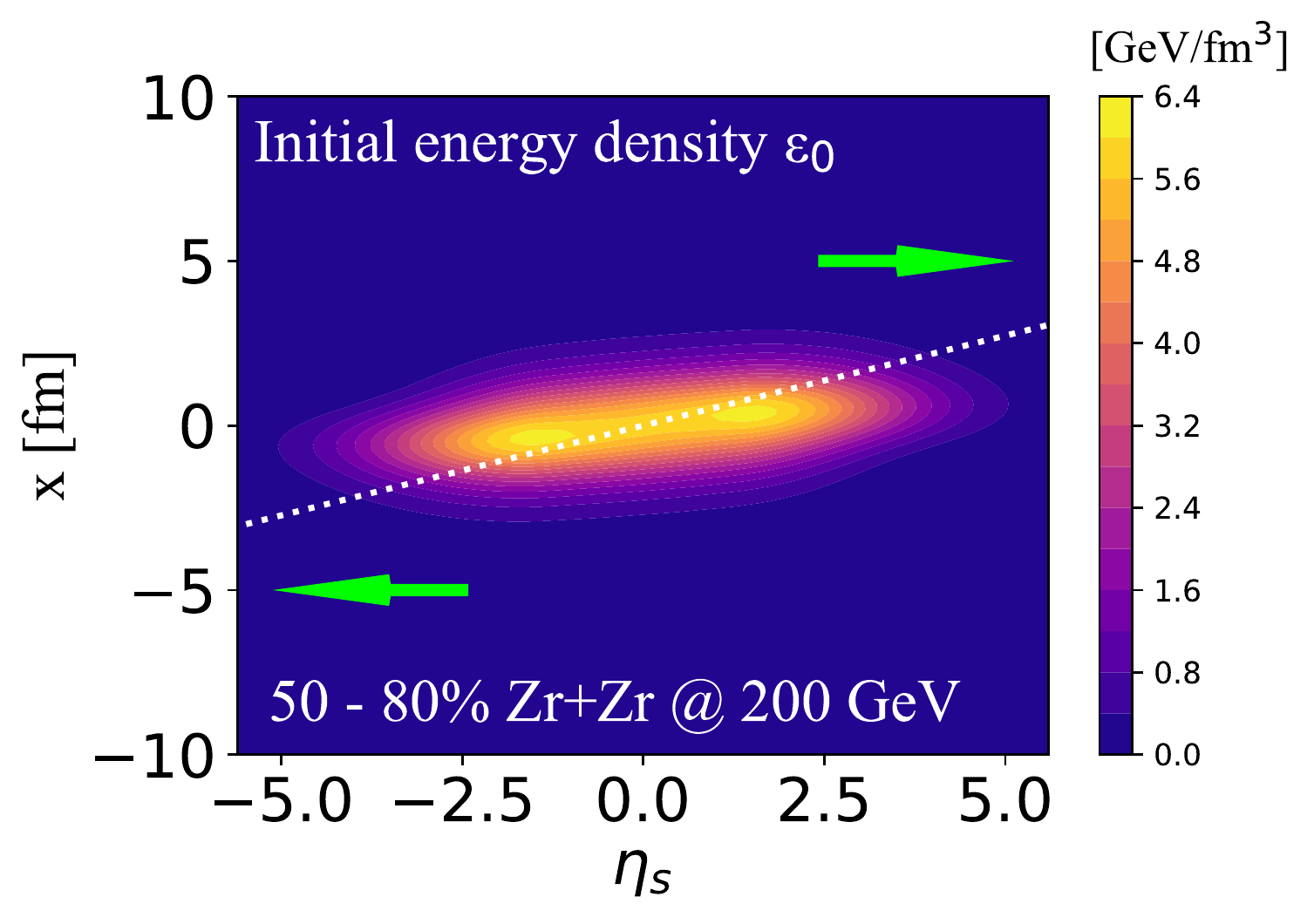}~
\end{center}
\caption{(Color online) The tilted initial energy density on the $\eta_{s}$-$x$ plane at $\tau_0$ = 0.6~fm/$c$ in 50-80\% Au+Au ($b$ = 11.4~fm), Ru+Ru ($b$ = 9.3~fm) and Zr+Zr ($b$ = 9.6~fm) collisions at $\sqrt{s_\text{NN}}=200$~GeV. The dotted line (snow color) shows the counter-clockwise tilted initial condition with respect to the $x=0$ axis in the $\eta_{s}$-$x$ plane, and the arrow (lime color) sketches the motion at forward and backward rapidity. }
\label{f:auau200ed}
\end{figure}

Using the tilted initial condition above, we first present the energy density profile for different nuclei (Au, Ru and Zr).
In Fig.~\ref{f:auau200ed}, we present the energy density profile for 50-80\% Au+Au, Ru+Ru and Zr+Zr collisions at $\snn = 200$~GeV in the $\eta_{s} - x$ plane.
Here the model parameter $H_{t} = 4.5$ is taken for the Au+Au collisions (top panel) and $H_{t} = 3.9$ for the Ru+Ru collisions (middle panel), while $H_{t}=3.1$ for Zr+Zr collisions (bottom panel). Here $H_{t}$ are extracted from experimental data to describe the directed flow $v_{1}$ of charged particles and protons/anti-protons.
From Fig.~\ref{f:auau200ed}, the energy density distribution for different nuclei is not only shifted in the forward/backward rapidity direction, it is also tilted counter-clockwise with respect to $x=0$~\cite{Bozek:2011ua,Jiang:2021foj}.

\begin{figure}[!tbp]
\begin{center}
\includegraphics[width=0.75\linewidth]{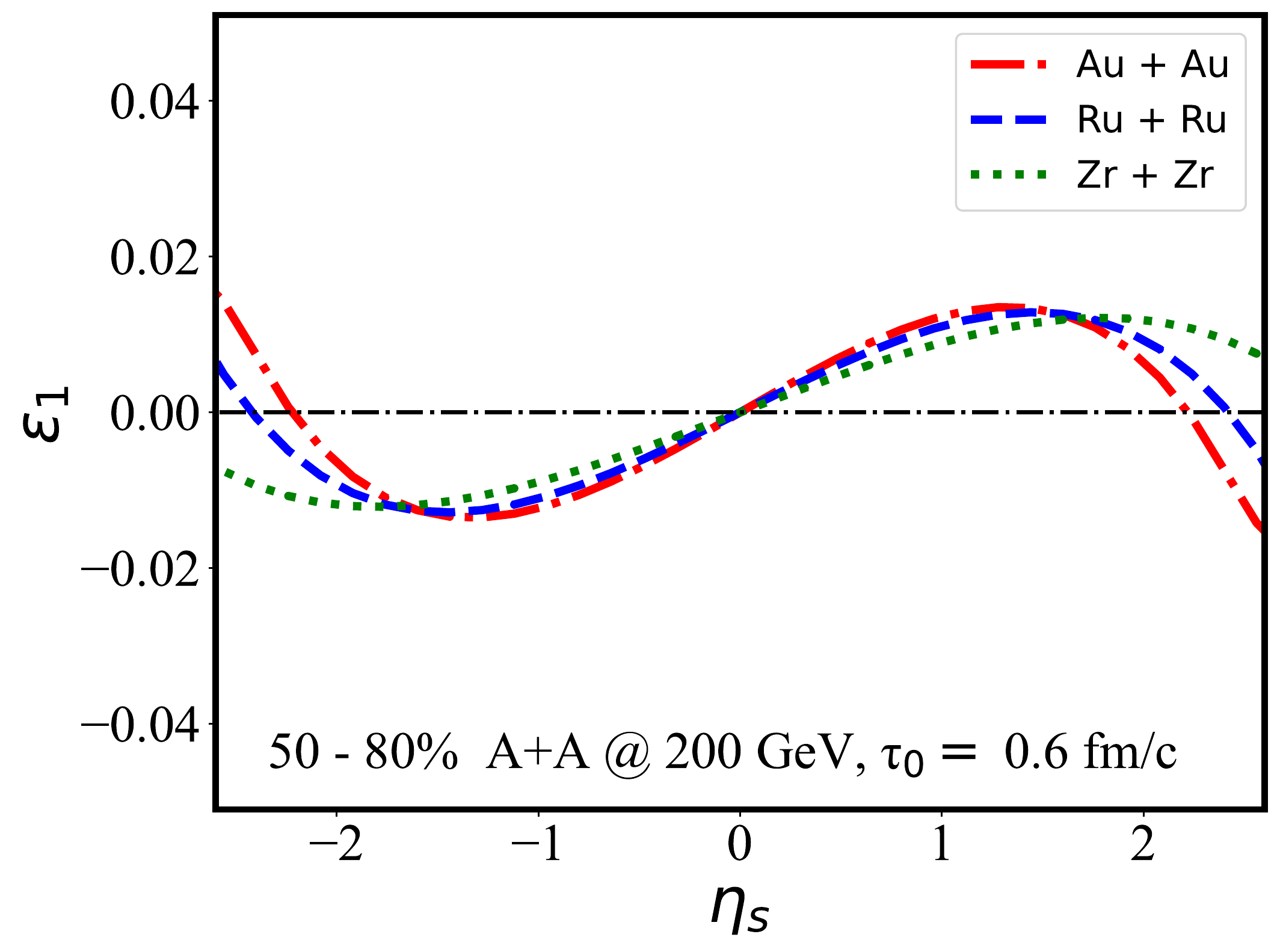}
\end{center}
\caption{(Color online) The first-order eccentricity coefficient $\varepsilon_{1}(\eta_{s})$ in 50-80\% nuclei-nuclei collisions at $\sqrt{s_\text{NN}}=200$ GeV.}
\label{f:auau200ecc1}
\end{figure}

To quantify the asymmetry strength of the initial energy density for different nuclei at initial state, we first present the first-order eccentricity coefficient $\varepsilon_1$ as a function of the space-time rapidity in Fig.~\ref{f:auau200ecc1}. The first-order eccentricity vector $\vec{\mathcal{E}}_{1}$ is defined as~\cite{Qiu:2011iv,Shen:2020jwv}:
\begin{equation}
\begin{aligned}
\vec{\mathcal{E}}_{1}\equiv\varepsilon_{1}(\eta_{s})e^{i\Psi_{1}(\eta_{s})}=
- \frac{\int d^{2}r \widetilde{r}^{3} e^{i\widetilde{\phi}}\varepsilon(r,\phi,\eta_{s})}{\int d^{2}r \widetilde{r}^{3} \varepsilon(r,\phi,\eta_{s})},
\label{eq:ecc1}
\end{aligned}
\end{equation}
where the angular distribution is calculated with respect to the energy density weighted center-of-mass point $(x_{0}(\eta_{s}), y_{0}(\eta_{s}))$ in every rapidity slice given by
\begin{equation}
\begin{aligned}
x_{0}(\eta_{s}) = \frac{\int d^{2} r x \varepsilon(r,\phi,\eta_{s})}{\int d^{2}r \varepsilon(r,\phi,\eta_{s})},
\label{eq:xy1}
\end{aligned}
\end{equation}
\begin{equation}
\begin{aligned}
y_{0}(\eta_{s}) = \frac{\int d^{2} r y \varepsilon(r,\phi,\eta_{s})}{\int d^{2}r \varepsilon(r,\phi,\eta_{s})},
\end{aligned}
\end{equation}
where $\widetilde{r}(x,y,\eta_{s})=\sqrt{(x-x_{0})^{2}+(y-y_{0})^{2}}$ is the transverse radius and $\widetilde{\phi}(x,y,\eta_{s})=\arctan[(y-y_{0})/(x-x_{0})]$ is the azimuthal angle, respectively. Please note that $\varepsilon_1$ in Eq.~(\ref{eq:ecc1}) presents the first-order eccentricity coefficient and $\Psi_1$ gives the corresponding participant plane angle. The $\vec{\mathcal{E}}_{1}$ with respect to $\eta_s$ quantifies the amount of tilt a fireball needs to produce the light hadrons $v_{1}$.

In Fig.~\ref{f:auau200ecc1}, we present the $\varepsilon_1$ as a function of space-time rapidity in 50-80\% Au+Au and isobar collisions at $\snn=200$ GeV. $\varepsilon_1$ is an odd function of space-time rapidity $\eta_s$ and  positive/negative in the $+$/$-\eta_s$ region.
The slopes of $\varepsilon_1$ for three nuclei are positive at mid-rapidity but flip sign within $|\eta_s| > 1.5$.
This will further affects the evolution of the nuclear medium in the hydrodynamic simulation.

\begin{figure}[!tbp]
\begin{center}
\includegraphics[width=0.75\linewidth]{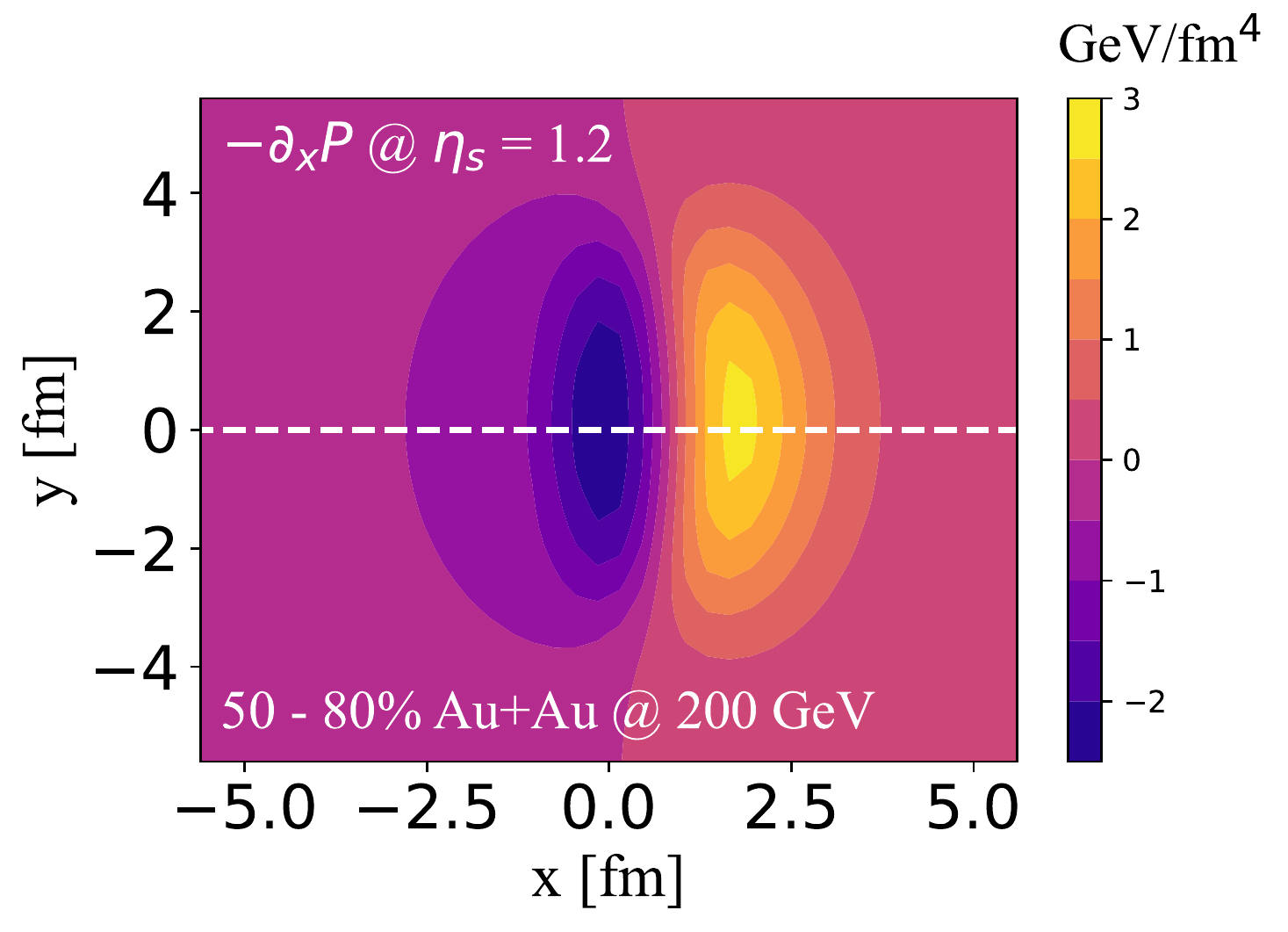}~\\
\includegraphics[width=0.75\linewidth]{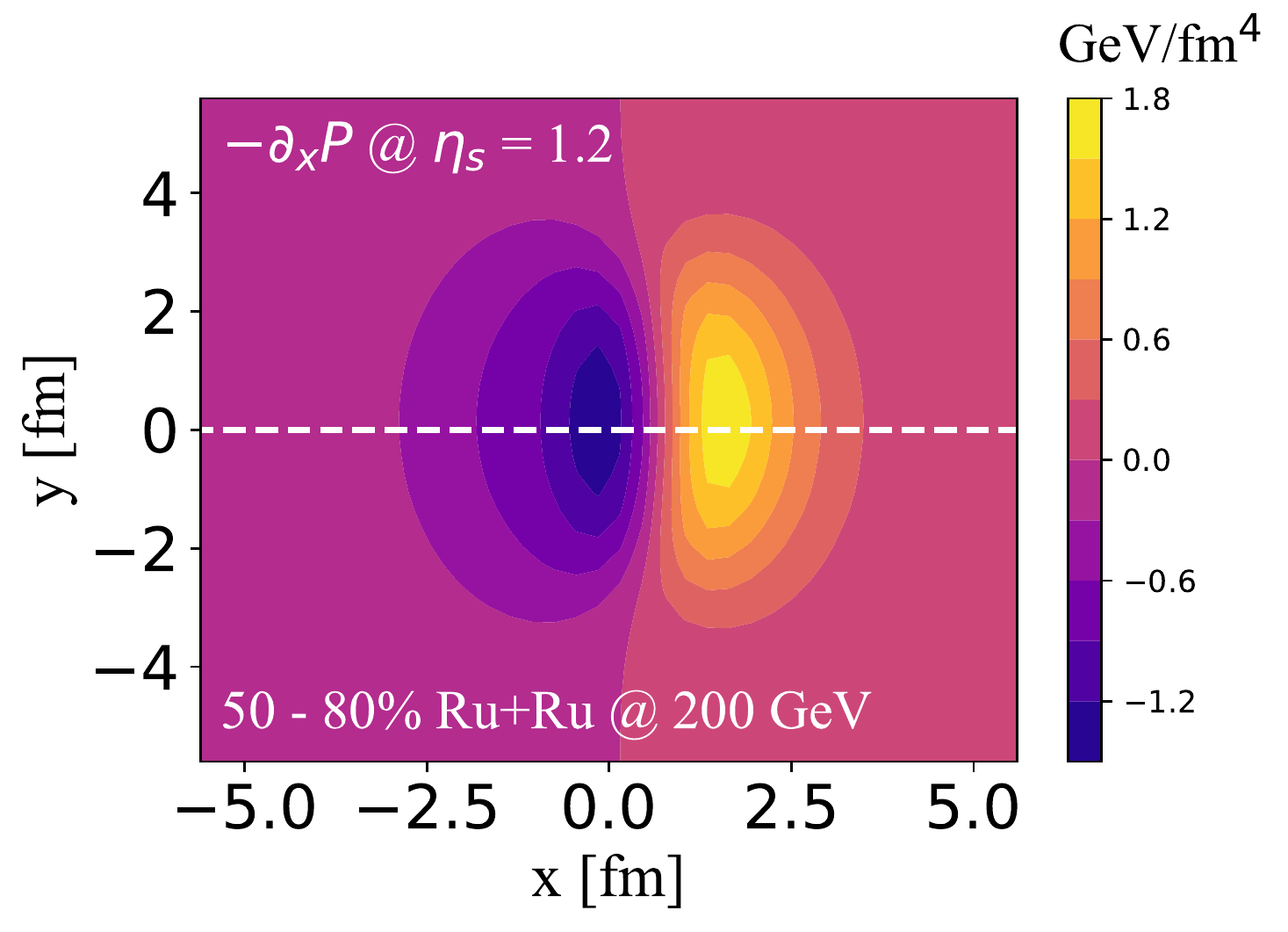}~\\
\includegraphics[width=0.75\linewidth]{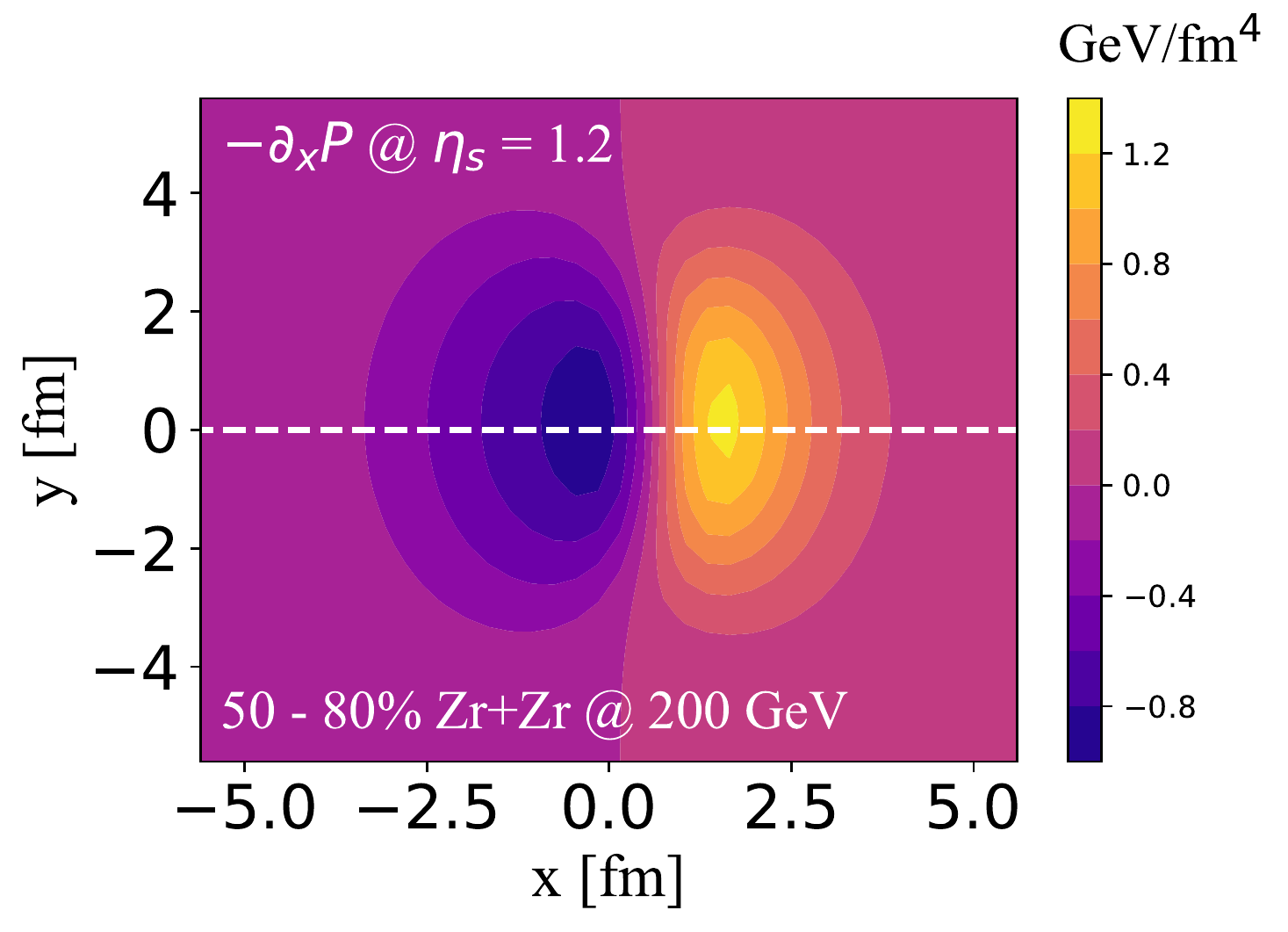}~
\end{center}
\caption{(Color online) The pressure gradient $-\partial_{x}P$ on the $x$-$y$ plane at initial proper time $\tau_0 = 0.6$~fm/$c$ in 50-80\% Au+Au and isobar collisions.}
\label{f:auau200pg}
\end{figure}

In addition to the initial energy density profile along the rapidity direction, we also present the initial pressure gradient $-\partial_{x}P$ in the transverse plane, which directly drives the radial flow of nuclear matter. In Fig.~\ref{f:auau200pg}, the initial $-\partial_{x}P$ distribution in the $x$-$y$ plane at a fixed forward rapidity $\eta_s=1.2$
is presented for nuclei Au, Ru and Zr, where the parameter values of $H_{t}$, $b$ are the same as used for Fig.~\ref{f:auau200ed}.
One may clearly find the positive/negative value of pressure gradient $-\partial_{x}P$ in the $+$/$-x$ direction that leads to the outward expansion of the nuclear medium.
From the top to bottom panels, we see that at $\eta_s=1.2$, the center (zero pressure) regions of these distributions are shifted towards $+x$ due to the counter-clockwise tilt of the initial energy density.
Whether the average $x$-component of the final-state hadron momentum will be positive or negative at a given rapidity depends on the average magnitude of $-\partial_{x}P$ in the corresponding transverse plane and how it evolves with time. This will be presented later in this work.

\subsection{Hydrodynamic evolution of the QGP}
We utilize the viscous hydrodynamic model CLVisc~\cite{Pang:2016igs,Pang:2018zzo,Wu:2018cpc,Chen:2017zte,He:2018gks}
to simulate the evolution of the QGP medium. The hydrodynamic equation satisfies~\cite{Jiang:2020big,Jiang:2018qxd,Denicol:2012cn,Romatschke:2009im,Romatschke:2017ejr}
\begin{equation}
\begin{aligned}
\partial_{\mu}T^{\mu\nu}=0,
\label{eq:tmn}
\end{aligned}
\end{equation}
where $T^{\mu\nu}$ is the energy-momentum tensor, and takes the following form:
\begin{equation}
\begin{aligned}
T^{\mu\nu}=\varepsilon u^{\mu}u^{\nu}-(P+\Pi)\Delta^{\mu\nu} + \pi^{\mu\nu}.
\label{eq:tensor}
\end{aligned}
\end{equation}
Here $\varepsilon$ is the energy density, $u^{\mu}$ is
the fluid four-velocity field, $P$ is the pressure, $\pi^{\mu\nu}$ is the shear stress tensor and $\Pi$ is the bulk pressure.
The projection tensor is defined as $\Delta^{\mu\nu} = g^{\mu\nu}-u^{\mu}u^{\nu}$, and the metric tensor $g^{\mu\nu} = \text{diag} (1,-1,-1,-1)$. In this work, we utilize the lattice QCD Equation of State (EoS) from the Wuppertal-Budapest group (2014)~\cite{Borsanyi:2013bia}, and the shear viscosity ratio is set as $\eta_{v}/s = 0.08$ ($\eta_{v}$ for the shear viscosity) for all collision systems. Following the recent studies~\cite{Jiang:2021foj,Jiang:2021ajc}, the bulk viscosity and net baryon density are ignored at this moment, and will also be taken into account in our future work~\cite{Zhao:2021vmu,Wu:2021fjf}.

When the local temperature of nuclear matter drops below the freeze-out temperature (we set $T_\text{frz}=137$~MeV)~\cite{Pang:2018zzo} , the Cooper-Frye mechanism~\cite{Cooper:1974mv} is used to calculate the spectra of hadrons on the freeze-out hypersurface.
Contributions from resonance decay are taken into account according to our previous work~\cite{Jiang:2021foj,Jiang:2021ajc}.

\subsection{Evolution of average pressure gradient and flow velocity with respect to proper time}

\begin{figure}[!tbp]
\begin{center}
\includegraphics[width=0.8\linewidth]{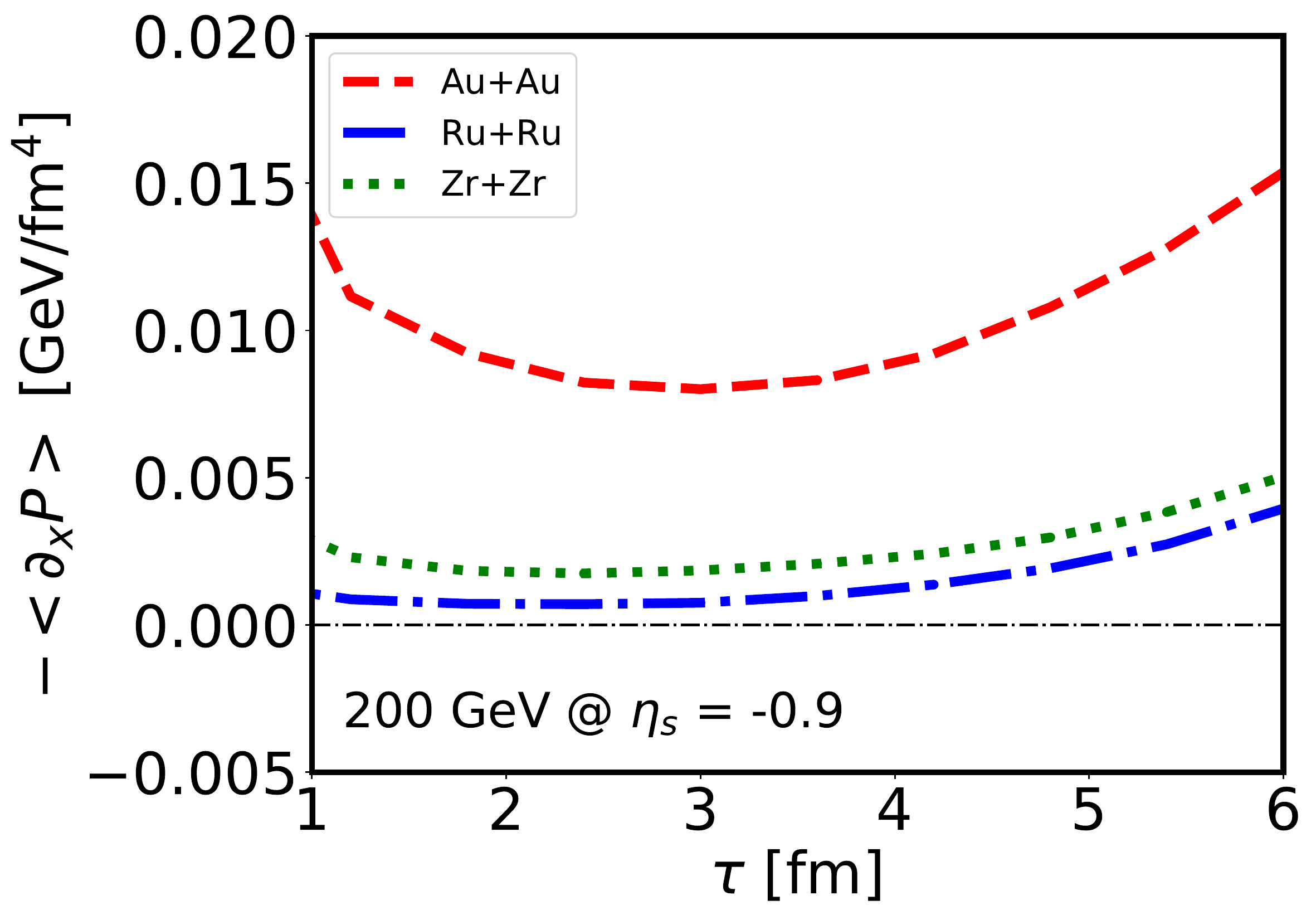}
\includegraphics[width=0.8\linewidth]{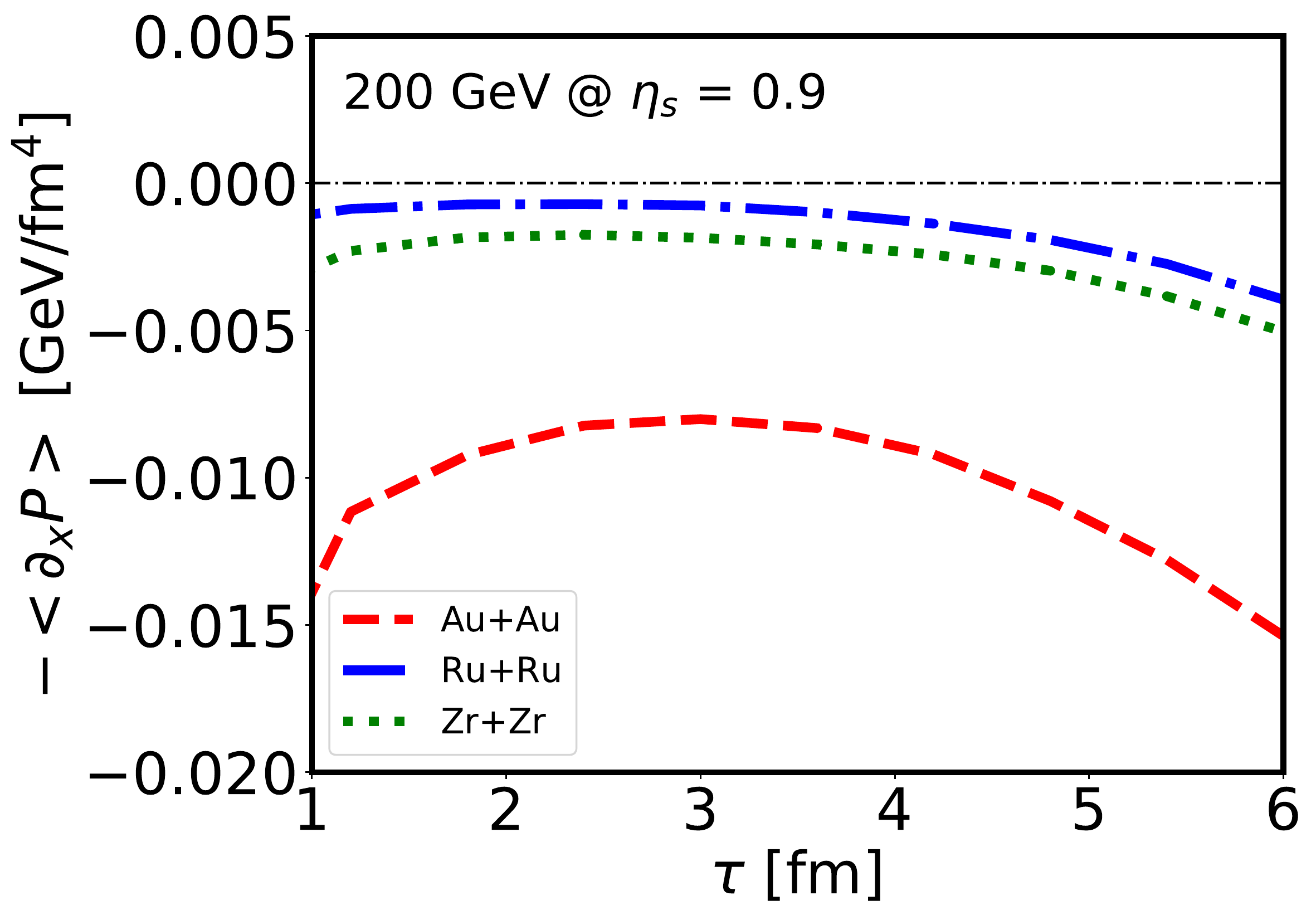}
\end{center}
\caption{(Color online) Time evolution of the average pressure gradient in the $x$ direction at $\eta_{s} = -0.9$ (upper panel) and $\eta_{s} = 0.9$ (lower panel) in 50-80\% Au+Au and isobar collisions at $\sqrt{s_\text{NN}}=200$~GeV.}
\label{f:auau200pge2}
\end{figure}

Hydrodynamic simulation presents how the imbalance of the initial energy density distribution is developed to the anisotropy of the final-state hadron momentum. In this subsection, we present how the fluid velocity $\langle v_{x} \rangle$ develops with respect to time at opposite rapidity $\eta_s$.
This will help us observe the development of directed flow $v_{1}$ and how it depends on the tilted initial geometry of the energy density.

As presented in Fig.~\ref{f:auau200pg} above, the tilted initial energy density leads to asymmetry of the pressure gradient along the $x$ direction.
In Fig.~\ref{f:auau200pge2}, we present how the average pressure gradient $-\langle\partial_{x}P \rangle$ evolves with time at a given $\eta_{s}$ ($\pm 0.9$ here) in 50-80\% Au+Au and isobar collisions at $\snn = 200$~GeV.
One may clearly observes that the evolution of $-\langle\partial_{x}P \rangle$ is significantly affected by the tilted initial energy density distribution.
The evolution of average pressure gradient -$\langle \partial_{x} P \rangle$ are anti-symmetric.~
One finds that the $-\langle\partial_{x}P \rangle$ remains positive with time at $\eta_s = -0.9$ and negative at $\eta_s = 0.9$, leading to a continuous force that accelerates the QGP medium outward expansion toward the $+x$ direction at $\eta_s = -0.9$ while toward the $-x$ direction at $\eta_s = 0.9$.
Little difference of the evolution of $-\langle\partial_{x}P \rangle$ is observed between the isobar Zr and Ru.
We note that unlike higher-order components of anisotropy, the first order eccentricity coefficient $\varepsilon_1$ and pressure gradient $-\langle\partial_{x}P \rangle$ are not necessarily positively correlated to each other.

\begin{figure}[!tbp]
\begin{center}
\includegraphics[width=0.75\linewidth]{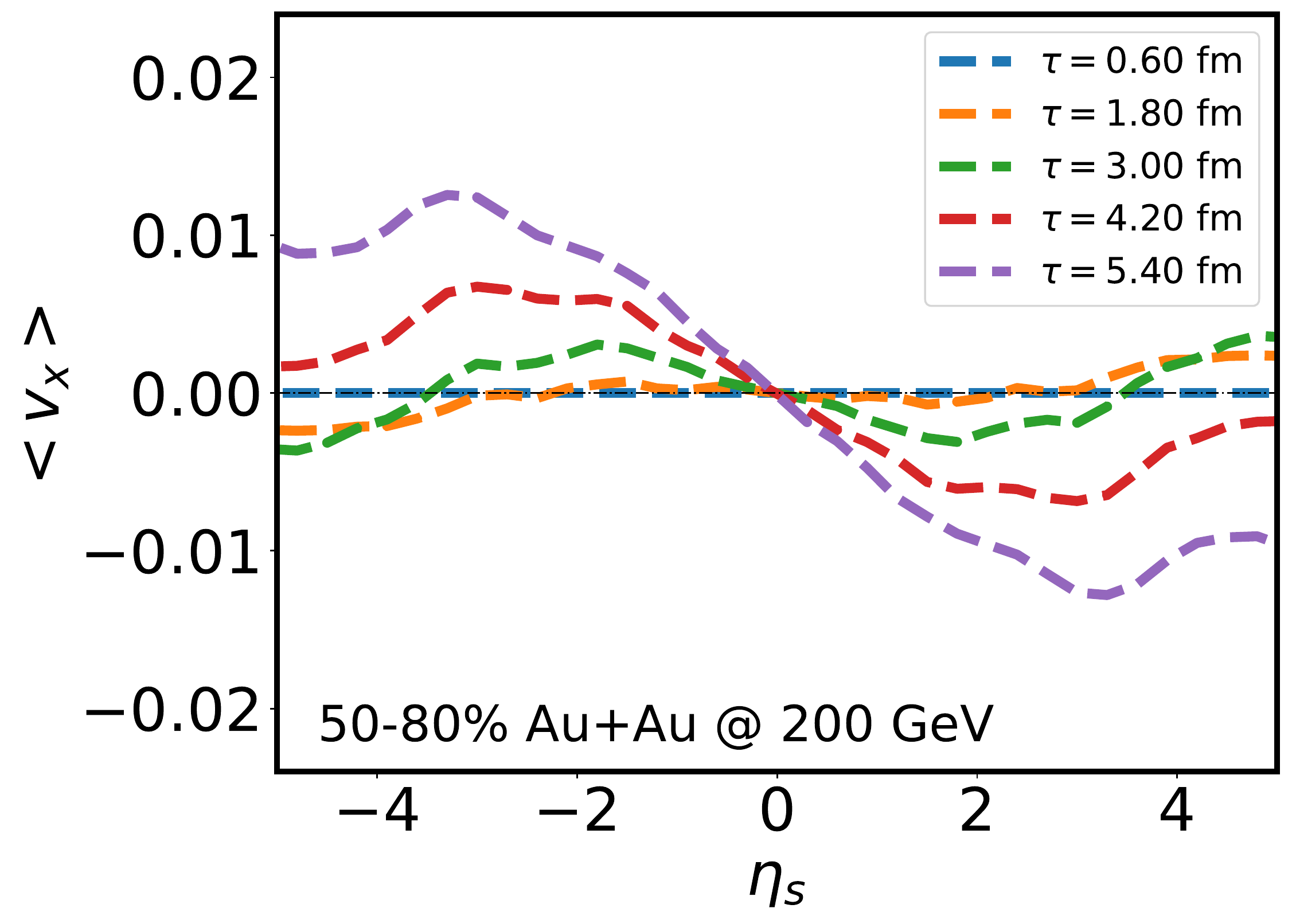}~\\
\includegraphics[width=0.75\linewidth]{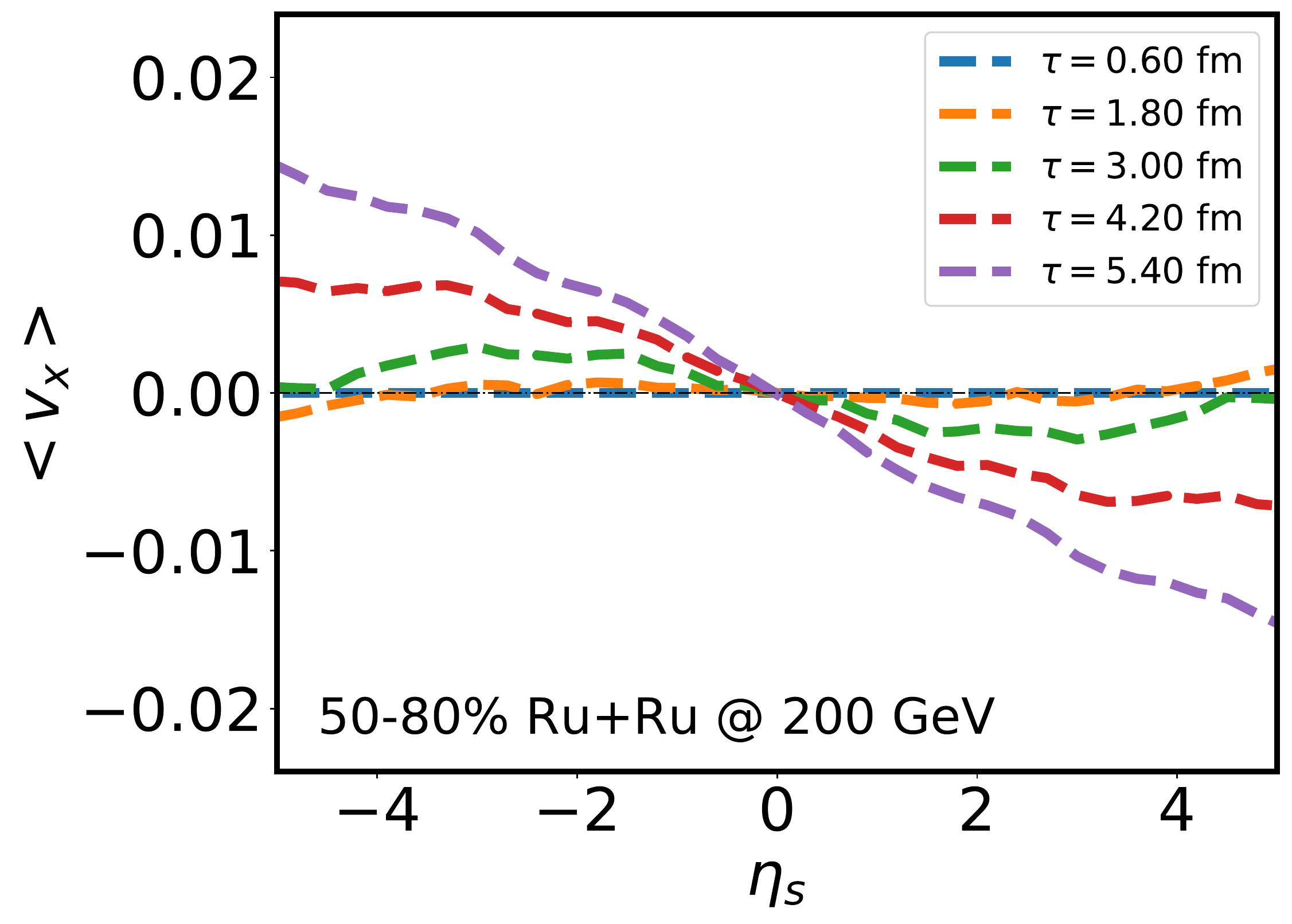}~\\
\includegraphics[width=0.75\linewidth]{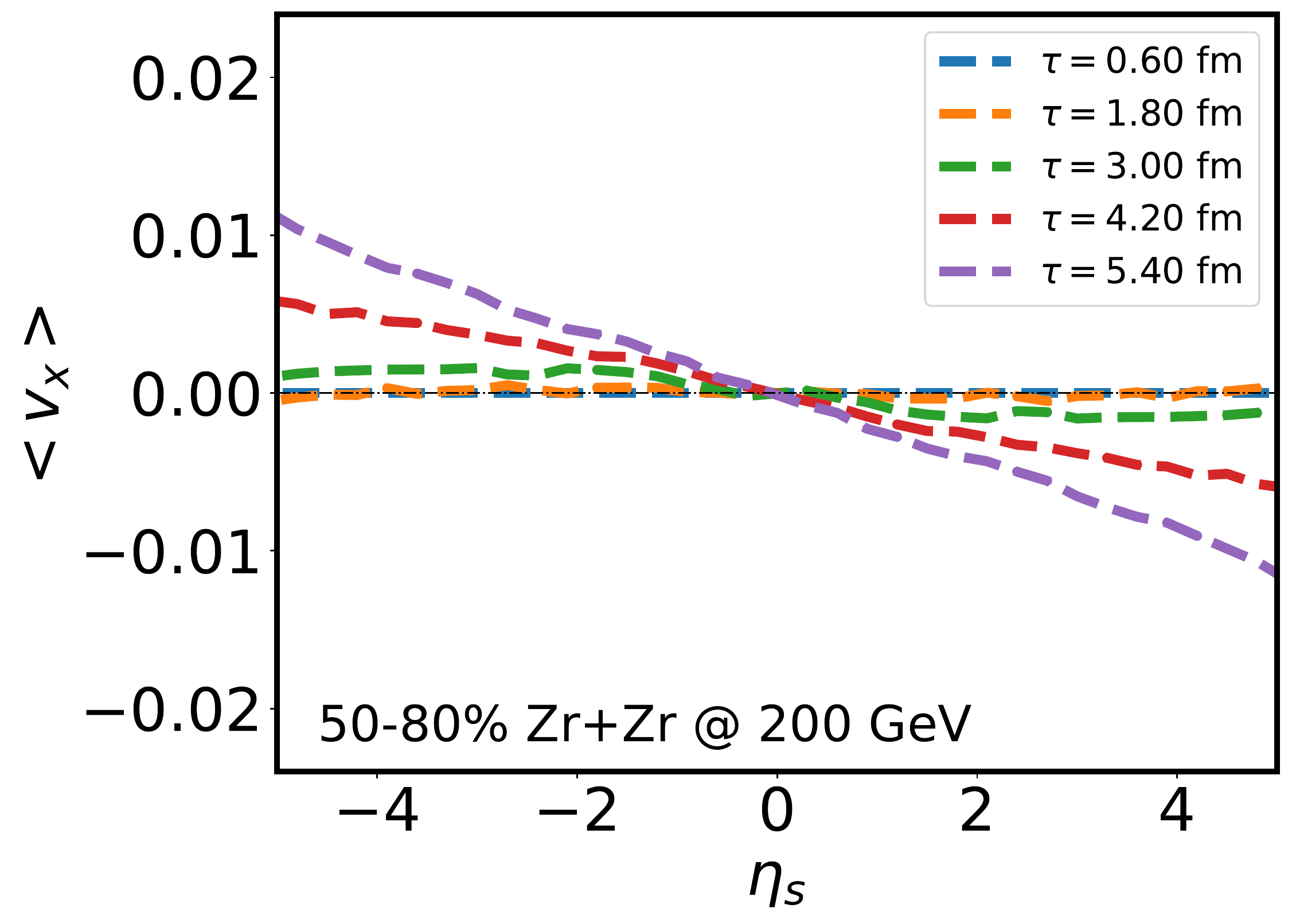}~
\end{center}
\caption{(Color online) Space-time rapidity dependence of the average flow velocity in the $x$ direction at different evolution times in 50-80\% Au+Au and isobar collisions at $\sqrt{s_\text{NN}}=200$~GeV.}
\label{f:auau200vx}
\end{figure}

Due to the tilted deformation of nuclear matter, medium expansion contributes to an overall force toward the $+$/$-x$ direction at backward/forward rapidity.
A direct outcome of such force is the asymmetric flow velocities in the corresponding direction.
In Fig.~\ref{f:auau200vx}, we present how the average flow velocity $\langle v_{x} \rangle$ develops with time.
The average flow velocity at a given proper time and space-time rapidity is defined as~\cite{Bozek:2010bi,Heinz:2013th}
\begin{equation}
\begin{aligned}
\langle v_{x}(\eta_{s}) \rangle = \frac{\int d^{2}r v_{x} \gamma \varepsilon(r,\phi,\eta_{s}) }{\int d^{2}r \gamma \varepsilon(r,\phi,\eta_{s})},
\label{eq:tar}
\end{aligned}
\end{equation}
where $\gamma=1/\sqrt{1-v_{x}^{2}-v_{y}^{2}-v_{\eta_{s}}^{2}}$ is the Lorentz boost factor.

In Fig.~\ref{f:auau200vx}, the average flow velocity $\langle v_x \rangle$ is positive/negative at backward/forward rapidity.
The magnitude of $\langle v_x \rangle$ increases with time due to the non-zero pressure gradient $-\langle\partial_{x}P \rangle$. One finds that a larger $|-\langle\partial_{x}P \rangle|$ leads to a larger $\langle v_x \rangle$ at $|\eta_s|$ around $|\eta_s|\approx 2$ regime than around $|\eta_s|\approx 1$ regime. The average flow velocity $\langle v_x \rangle$ here will directly produce the directed flow of the light hadrons.

\section{Numerical results}
\label{v1section3}

In this section, we present the numerical results for light hadrons yield and directed flow in Au+Au and isobar collisions at $\snn=200$ GeV using the above tilted initial condition and hydrodynamic model CLVisc.
In particular, we investigate how the directed flow depends on the effect of nuclear structure of the isobar Ru and Zr.

\begin{figure}[!tbp]
\begin{center}
\includegraphics[width=0.75\linewidth]{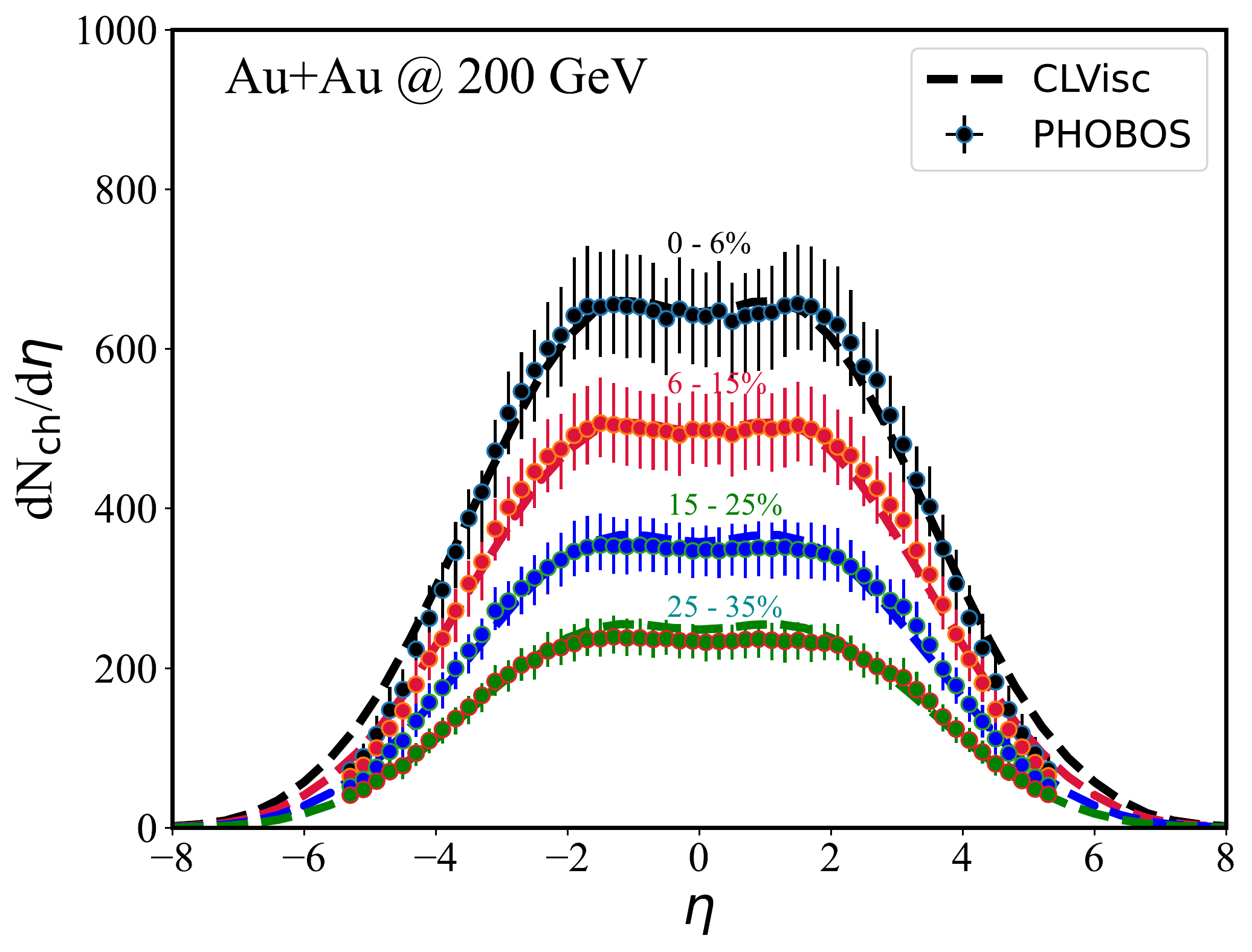}~\\
\includegraphics[width=0.75\linewidth]{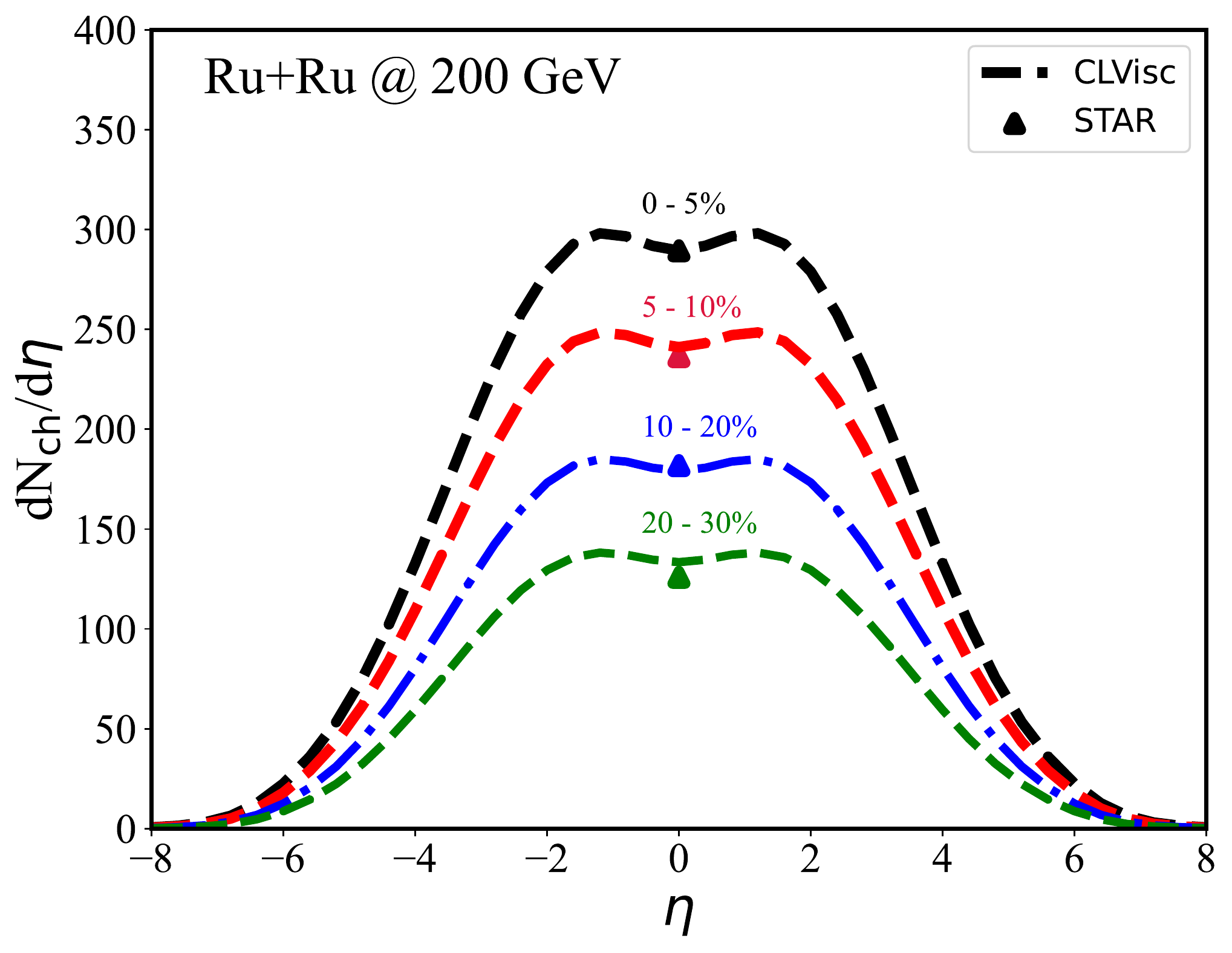}~\\
\includegraphics[width=0.75\linewidth]{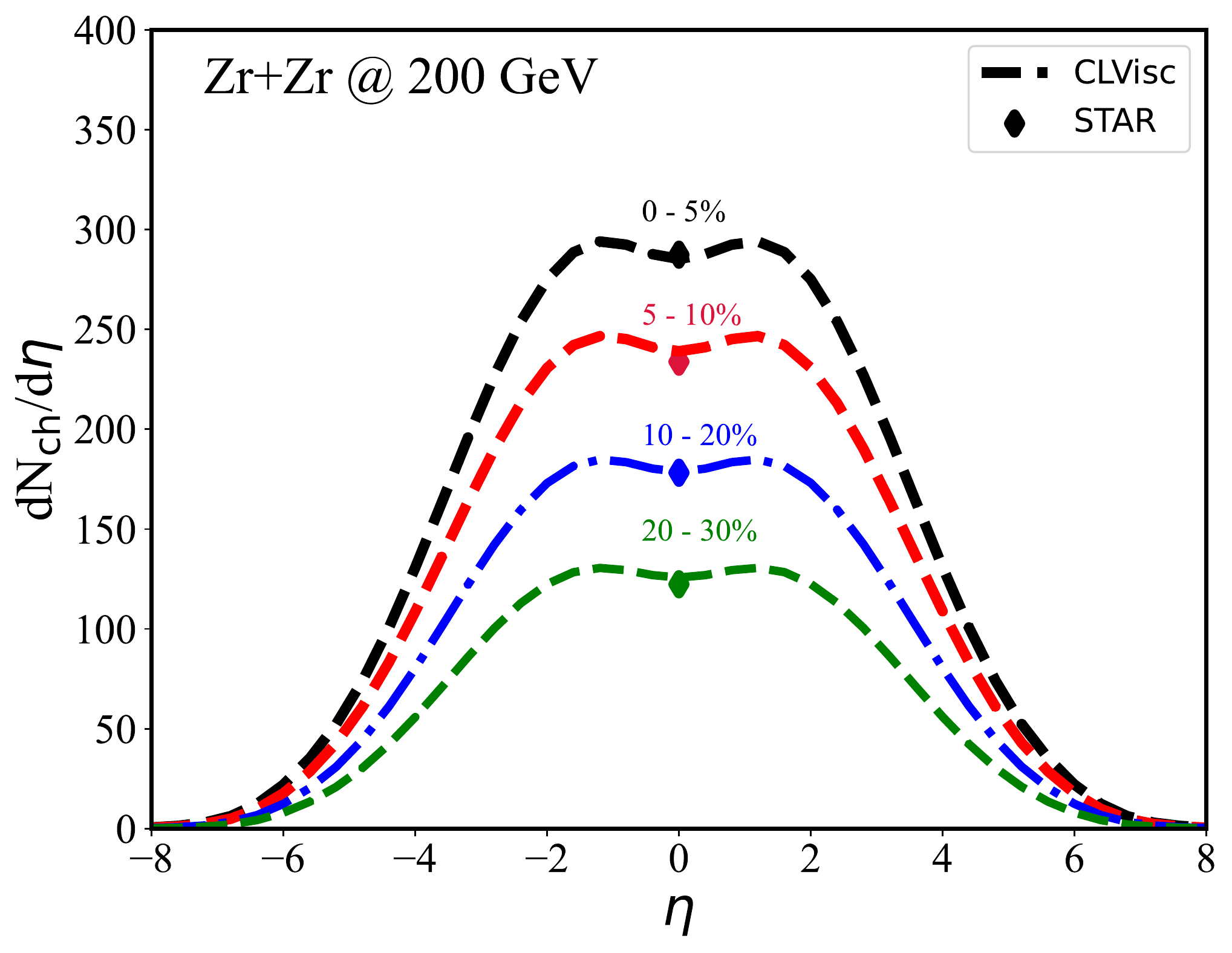}~
\end{center}
\caption{(Color online) Pseudorapidity distribution of charged light hadrons in Au+Au and isobar collisions at $\snn = 200$~GeV, compared between the CLVisc hydrodynamic calculation with three nuclei and the PHOBOS and STAR data~\cite{Alver:2010ck,STAR:2021mii}.}
\label{f:auau200dndeta}
\end{figure}

In Fig~\ref{f:auau200dndeta}, we show the pseudorapidity distributions of the charged particles in Au+Au, Ru+Ru and Zr+Zr collisions at $\sqrt{s_\text{NN}}=200$~GeV.
As discussed in Sec.~\ref{v1subsect2}, the hydrodynamic model parameters summarized in Tab.~\ref{t:modelparameters}
are adjusted to describe the charged light hadrons distributions in the most central collisions.
As shown in the figures, our calculation presents reasonable descriptions of the PHOBOS data on the $dN_\text{ch}/d\eta$ distributions for Au+Au collisions in serval centralities at $\snn=200$ GeV.
In addition, we note that the parameter $H_{t}$ only affects the deformation of the medium geometry,
but has a very weak impact on the $dN_\text{ch}/d\eta$ distributions~\cite{Jiang:2021ajc}.
This provides a reliable baseline for our further investigation of the light hadrons directed flow.
Please note we define centrality bins utilizing impact parameter, and the contribution from the centrality fluctuation is not included for current study.
In the future, it is necessary to use the multiplicity distribution with the given experimental data to characterize the centrality bins.

\begin{figure}[!tbp]
\begin{center}
\includegraphics[width=0.75\linewidth]{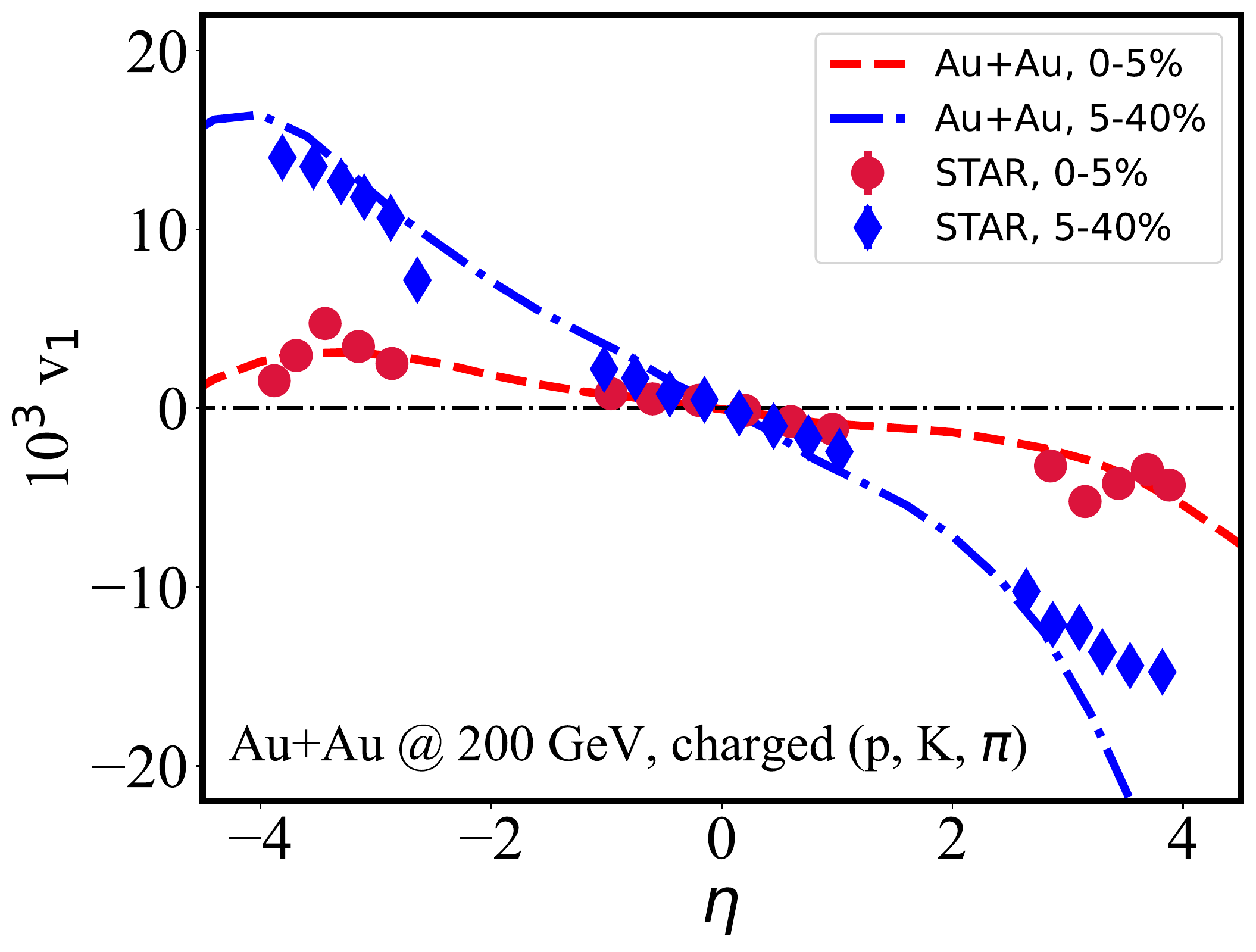}~\\
\includegraphics[width=0.75\linewidth]{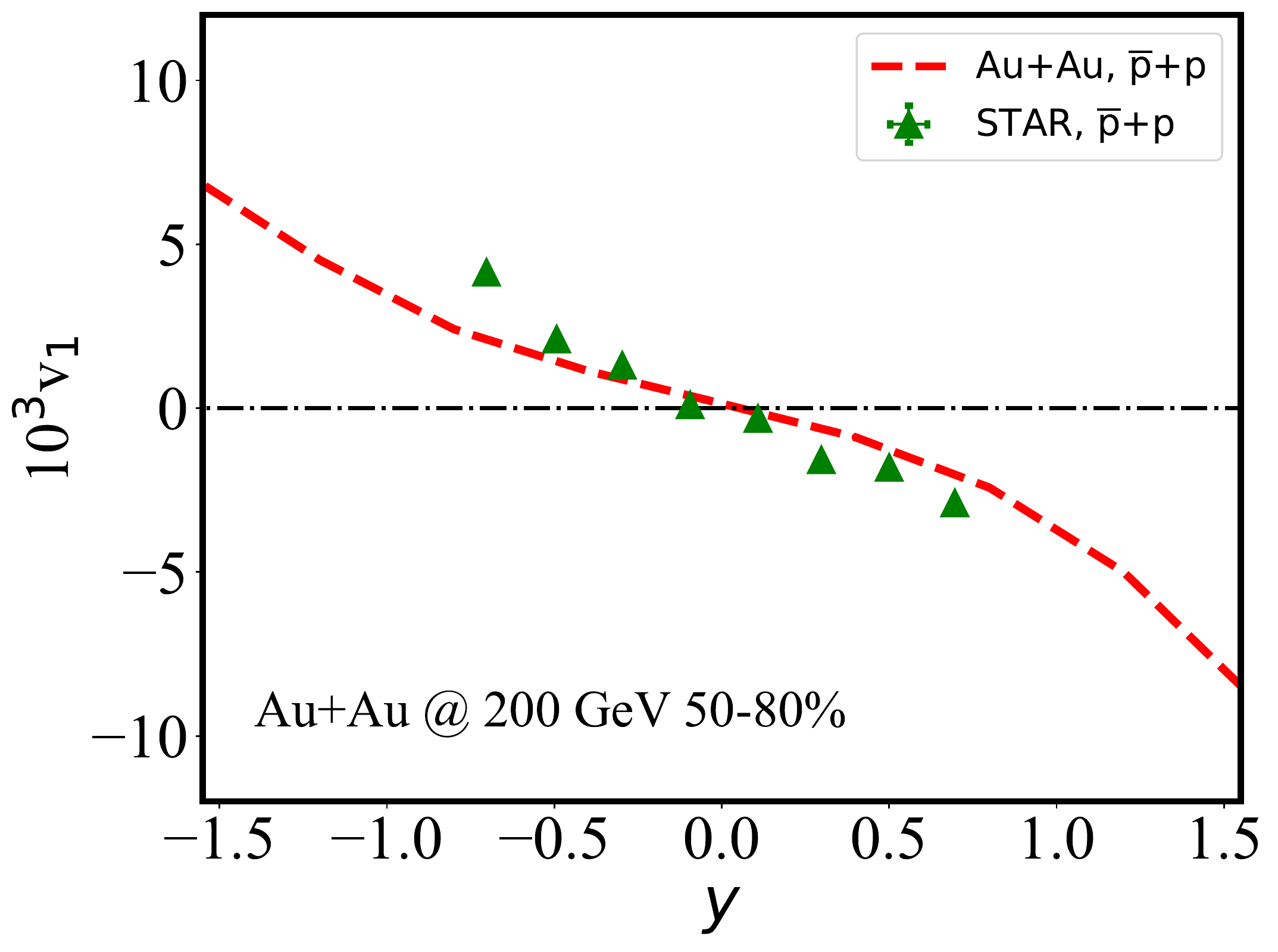}~
\end{center}
\caption{(Color online) Pseudorapidity and rapidity dependence of the directed flow of identified hadrons in 0-5\%, 5-40\% (upper panel) and 50-80\% (lower panel) Au+Au collisions at $\snn = 200$~GeV, compared between the CLVisc hydrodynamic calculation with the STAR data~\cite{Abelev:2008jga,qm2022ruzr}.}
\label{f:v1-200}
\end{figure}
\begin{figure}[!tbp]
\begin{center}
\includegraphics[width=0.75\linewidth]{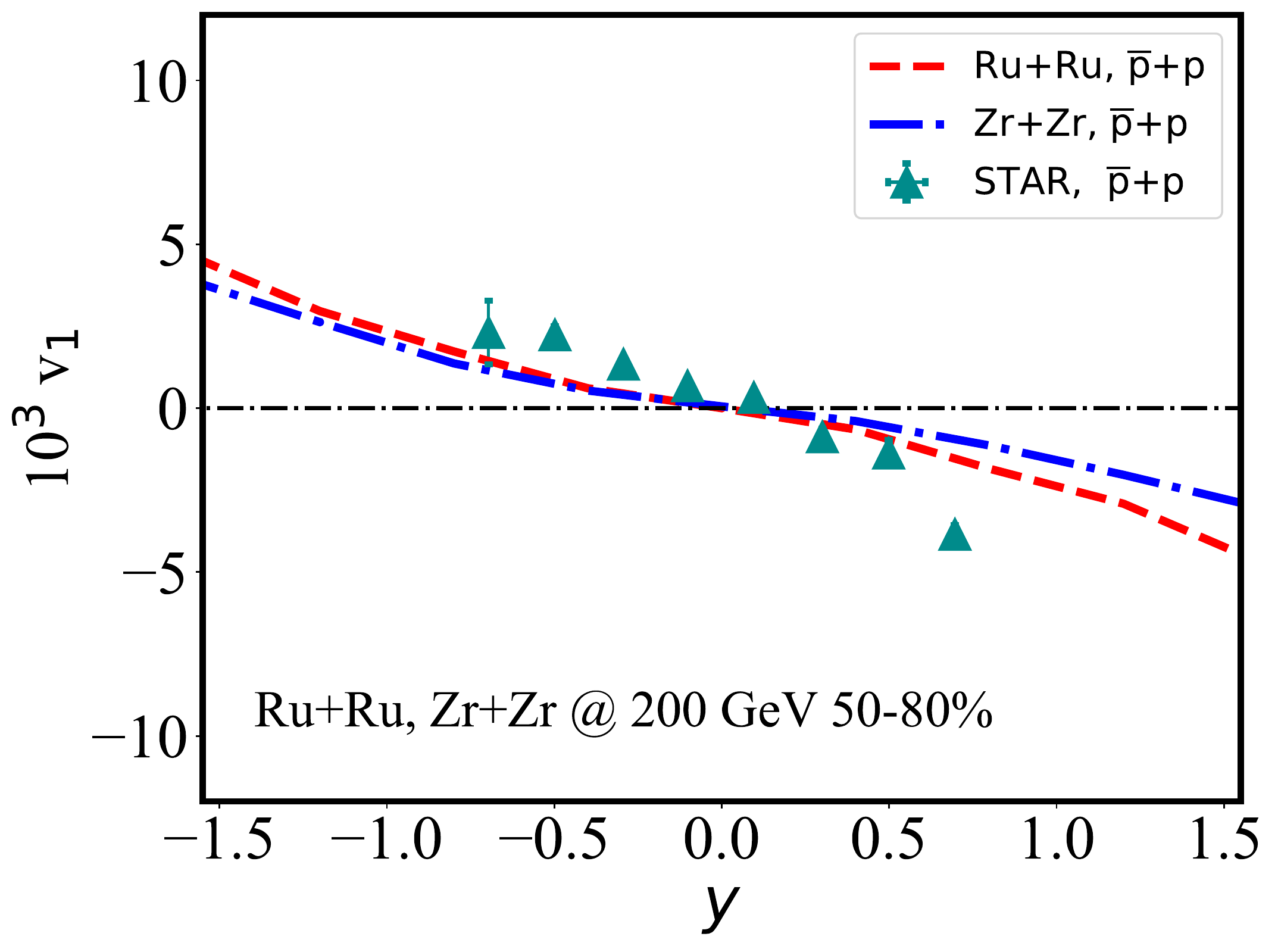}~
\end{center}
\caption{(Color online) Rapidity dependence of the directed flow coefficient in 50-80\% Ru+Ru and Zr+Zr collisions at $\snn = 200$~GeV, compared between the CLVisc hydrodynamic calculation with Case-1 nuclear structure and the STAR data~\cite{qm2022ruzr}.}
\label{f:v1-zrru}
\end{figure}

We then present the identified particle directed flow $v_1$ as a function of pseudorapidity.
Following our previous work~\cite{Jiang:2021ajc,Jiang:2021foj}, here $v_{1}(\eta)$ is calculated via
\begin{equation}
\begin{aligned}
v_{1}(\eta)=\langle\cos(\phi-\Psi_{1})\rangle=\frac{\int\cos(\phi-\Psi_{1})\frac{dN}{d\eta d\phi}d\phi}{\int\frac{dN}{d\eta d\phi}d\phi},
\label{eq:v1}
\end{aligned}
\end{equation}
where $\Psi_{1}$ is the first-order event plane of the collision~\cite{Bozek:2010bi}.
The directed flow $v_1$ is analyzed with soft hadrons within $0 < p_{\textrm{T}} < 3.0$ GeV in our current hydrodynamic model. Besides, since the experiment usually uses $0.2 < p_{\textrm{T}} < 2.0$ GeV for charge particles in the analysis, we will use such $p_{\textrm{T}}$ range in future study.
In this work, we use the optical Glauber model to construct the initial energy density distribution of the nuclear matter,
and the initial event-by-event fluctuations are neglected~\cite{Bozek:2010bi,Jiang:2021foj}.
As a result, the event plane here is the same as the spectator plane determined using the deflected neutrons in realistic experimental measurements.
The Monte Carlo Glauber model contributes the effect of event-by-event fluctuations, which would enhances the directed flow at large forward and backward rapidity.
A more consistent study will be conducted in our future work after event-by-event fluctuations are taken into account.

Using above setups, we show the light hadrons $v_1$ in Au+Au collisions for different centrality classes at $\snn = 200$~GeV in Fig.~\ref{f:v1-200}, upper panel for charged particles and lower for proton ($p$) and anti-proton ($\bar{p}$). Our calculations for charged particle $v_1$ within $-4.5 < \eta < 4.5$ are consistent with the STAR data. As expected, the distribution of the identified particles $v_1$ is consistent with that of the average flow velocity $v_x$ of the nuclear matter.

For isobar collisions, as illustrated in Fig.~\ref{f:v1-zrru}, within our hydrodynamic framework, we are able to describe the directed flow $v_1$ of protons ($p^{\pm}$) at either Ru+Ru or Zr+Zr by adjusting the $H_{t}$ parameter.
The difference of $v_1$ value between the two nuclei is less than 0.001 within $|y|<1.5$.
If one decreases the value of $H_{t}$, the slope of $v_1$ {\it vs.} $y$ will decrease near the mid-rapidity region and further deviate from the experimental data.
This implies the importance of the tilted initial energy density distribution in understanding the rapidity dependence of light hadrons $v_1$ observed in experiments.
Based on Figs.~\ref{f:v1-200}-\ref{f:v1-zrru}, we find with a different $H_t$ for isobar and Au+Au collisions at same centrality bins, a clear system size dependence of directed flow is observed.
Smaller nuclei has a smaller $H_{t}$ and leads to a smaller directed flow at mid-rapidity.

\begin{figure}[!tbp]
\begin{center}
\includegraphics[width=0.75\linewidth]{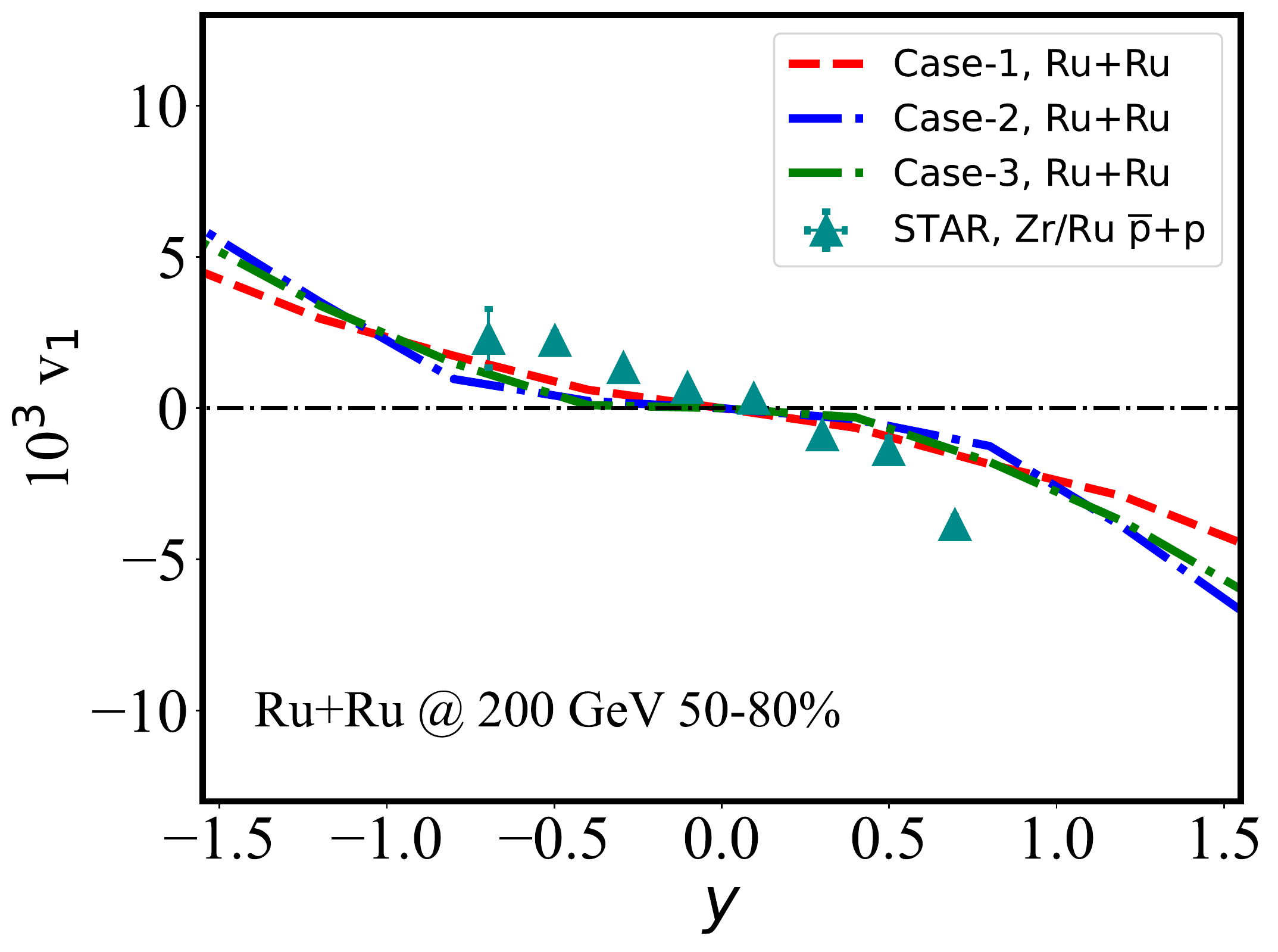}~\\
\includegraphics[width=0.75\linewidth]{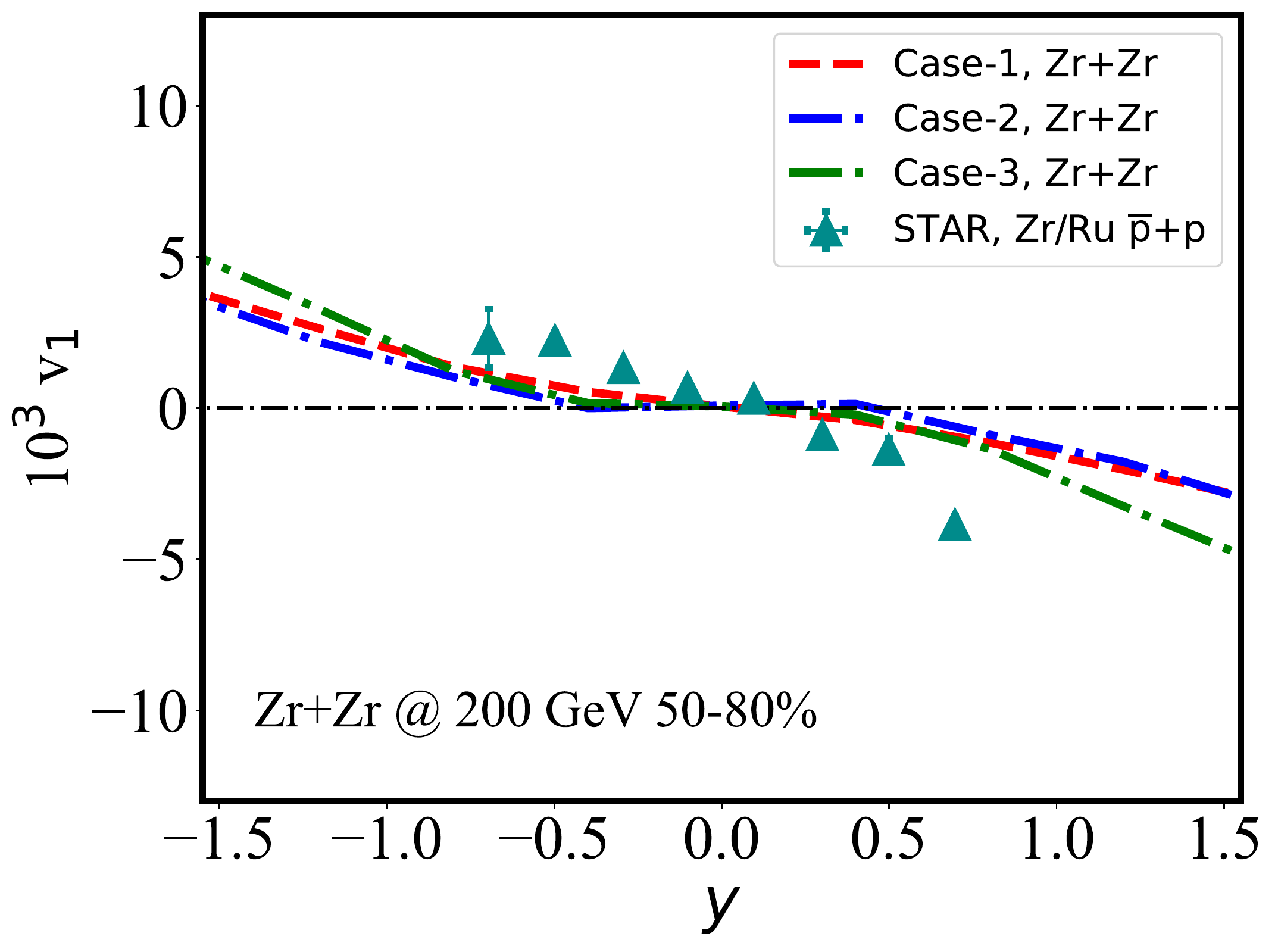}~
\end{center}
\caption{(Color online) Rapidity dependence of the directed flow coefficient in 50-80\% Ru+Ru collisions (upper panel) and Zr+Zr collisions (lower panel) at $\snn = 200$~GeV, compared between the CLVisc hydrodynamic calculation with three nuclear structure setups and the STAR data~\cite{qm2022ruzr}.}
\label{f:v1-2760}
\end{figure}

\begin{table}[h!]
    \centering
    \caption{\label{t:para2} Nuclear structure parameters for Ru and Zr from Ref.~\cite{STAR:2021mii}.}
    \begin{tabular}{ |c|c|c| c |c| c| c | }
    \hline
          &\multicolumn{2}{c|} {Case-1} & \multicolumn{2}{c|} { Case-2}  & \multicolumn{2}{c|} { Case-3}         \\
    \hline
    Parameters         & Ru+Ru   & Zr+Zr  & Ru+Ru   & Zr+Zr     & Ru+Ru       & Zr+Zr        \\
    \hline
    $R$~(fm) & 5.067& 4.965    & 5.085        & 5.02   & 5.085 & 5.02            \\
    \hline
    $d$~(fm) & 0.500 &0.556     & 0.46       & 0.46    & 0.46 & 0.46              \\
    \hline
    $\beta_{2}$ &0.0 &0.0         & 0.158       & 0.08    & 0.053  & 0.217        \\
    \hline
    \end{tabular}
\end{table}

So far, we have focused on the discussion of spherical nuclei (from DFT) in which the quadruple deformity parameter $\beta_{2}=0$ (as listed in Tab.~\ref{t:parameters}).
As pointed out in Refs.~\cite{Nijs:2021kvn,Zhang:2021kxj,Jia:2021oyt,Xu:2021uar,Li:2022bhl,Zhao:2022grq,Ma:2022dbh,Jia:2022qrq}, the elliptic flow in the most central collisions is sensitive to nuclear deformation, as deformed nuclei colliding at impact parameter $b=0$
can induce a large eccentricity on the collision orientation. In order to study the effect of nuclear deformation on light hadrons $v_{1}$, within our hydrodynamic framework, the extended WS parameters of nuclei listed in Tab.~\ref{t:para2} are utilized.
The two sets (Case-2 and Case-3) have the same $R$ and $a$ parameters and different deformations which are constrained by $e+A$ scattering experiments and calculations based on a finite-range droplet macroscopic model~\cite{Raman:2001nnq,Pritychenko:2013gwa}
and the folded-Yukawa single-particle microscopic model~\cite{Moller:1993ed}.
Because of the additional protons in Ru, the charge radius of Ru is larger than that of Zr.
Other parameters during the QGP evolution are the same as the spherical nuclei (Case-1).

Fig.~\ref{f:v1-2760} shows the protons $v_{1}$ in Ru+Ru/Zr+Zr collisions at $\snn=200$ GeV with various combinations of Ru and Zr deformities.
The comparison between the three nuclear structure shows that the $\beta_{2}$ deformations seems not essential to change the $v_{1}$ slope and magnitude at central pseudorapidity ($|y|<0.5$).
We see that the slope of $v_{1}$ changes insignificantly due to the finite $\beta_{2}$ between Case-2 and Case-3 for both Ru+Ru and Zr+Zr collisions.
The impact of different parameters $R$, $d$ and $\beta_{2}$ on protons $v_{1}$ implies that the nuclear structure with quadrupole deformation only
slightly affects the directed flow of the final state particles.
We would note here that the $v_{1}$ between Ru and Zr at large rapidity may be sensitive to the shape of nuclei.
Besides, many low energy experiments show that Zr has a large octupole deformation corresponding to a large value of $\beta_{3}$.
Theory studies find the average of first-order eccentricity coefficient $\varepsilon_1$ is proportional to the average of $\beta_{3}^{2}$,
and the octupole deformation should affects the $v_{1}$ of soft hadrons~\cite{Jia:2021tzt,Zhang:2021kxj}.
The effect of octupole deformation will be investigated in an upcoming work.

We note that since we use the optical Glauber model (smooth initial condition), our calculation is restricted to the rapidity-odd component of light hadrons $v_1$ here. The rapidity-even component $v_1^{\textrm{even}}$, including its non-trivial $p_\mathrm{T}$ dependence~\cite{Teaney:2010vd,Luzum:2010fb,Gale:2012rq} is beyond the scope of this study and will be investigated in the future.

\section{Summary}
\label{v1section4}

In this work, we present a systematic study on how the initial medium profile of nuclei evolves to the light hadrons directed flow in heavy ion collisions.
Three different nucleus-nucleus collisions --Au+Au, Ru+Ru and Zr+Zr-- are compared for the tilted initial energy density distribution, and their subsequent space-time evolutions are simulated utilizing the hydrodynamic model CLVisc.

Using our (3+1)-D hydrodynamic model CLVisc, we have calculated the directed flow $v_{1}$ for light hadron as a function of (pseudo-)rapidity and centrality for Au+Au and isobar collisions at $\snn=200$ GeV.
Our model can give a good description of light hadrons $v_1$ in central and peripheral Au+Au and isobar collisions measured by the STAR Collaborations.
Our results show that light hadrons $v_{1}$ have a strong system size dependence, i.e., $v_1$ is larger for small systems, due to weaker nuclear stopping effect.
The system size dependence is also seen from the centrality dependence:
$v_1$ becomes large when moving from central to peripheral collisions, which is clearly observed in Au+Au collisions and can be tested by future experiments.
We further find that the Au+Au and isobar semi-central collisions generate an imbalance between forward/backward moving nucleus,
induce a counter-clockwise tilt of the initial medium profile in the $x$-$\eta_s$ plane,
and generate a non-zero average pressure gradient $-\langle \partial_x P\rangle$ with respect to time at backward/forward rapidity.
A comparison to the RHIC-STAR data indicates that the tilted initial energy density profile (or fireball geometry)
is an essential factor to generate the observed light hadrons $v_1$ in Au+Au and isobar collisions at $\snn=200$ GeV.
We finally find the effect of nuclear structure with quadrupole deformation insignificantly affects the light hadron directed flow at mid-rapidity.

Our study provides a step forward in understanding of the origin of the light hadrons directed flow in Au+Au and isobar collisions.
However, in addition to the effect of tilted initial energy density, other sources also contribute to the size and sign of directed flow.
For example, (1) The extremely strong electromagnetic field produced in the non-central nucleus-nucleus collisions results in directional drift of charged quarks ($u,d,s$) and influences the charged particle $v_1$~\cite{Inghirami:2019mkc,Gursoy:2014aka,Gursoy:2018yai}. Although this effect is suggested to be smaller than the effect of initial titled geometry, it is important to understand the splitting of $v_{1}$ for identified light hadrons in isobar collisions~\cite{Inghirami:2019mkc}. (2) The fluid velocity field could provide an additional contribution to the light hadrons directed flow~\cite{Ryu:2021lnx}. In particular, they could affect the initial baryon density distribution and thus the nuclear matter properties~\cite{Bozek:2010bi}.
(3) The light hadrons $v_1$ can also be affected by the nuclear stopping effect and hadronic cascade after the QGP evolution, especially at lower collision energy~\cite{Shen:2020jwv,Ryu:2021lnx,Bozek:2022svy}.
(4) The light hadrons elliptic flow $v_{2}$ and triangular flow $v_{3}$ could set more constraints on the nuclear structure, but we will not discuss it here because we limited ourselves to a smooth initial condition without the contribution of event-by-event fluctuations.
These should be investigated in our future study for a more exactly understanding of the directed flow.

\begin{acknowledgements}
We are grateful for helpful discussions with Xiangyu Wu and Shanshan Cao.
This work was supported by the National Natural Science Foundation of China (NSFC) under Grant Nos.~11935007,
Guangdong Major Project of Basic and Applied Basic Research No.~2020B0301030008, the Natural Science Foundation of Hubei Province No.~2021CFB272, the Education Department of Hubei Province of China with Young Talents Project No.~Q20212703,
the Open Foundation of Key Laboratory of Quark and Lepton Physics (MOE) No.~QLPL202104 and
the Xiaogan Natural Science Foundation under Grant No.~XGKJ2021010016.
\end{acknowledgements}

\bibliographystyle{unsrt}
\bibliography{heavy_v1ref}

\begin{thebibliography}{100}

\bibitem{PHENIX:2003qra}
S.~S. Adler et~al.
\newblock {Elliptic flow of identified hadrons in Au+Au collisions at
  s(NN)**(1/2) = 200-GeV}.
\newblock {\em Phys. Rev. Lett.}, 91:182301, 2003.

\bibitem{ALICE:2010suc}
K~Aamodt et~al.
\newblock {Elliptic flow of charged particles in Pb-Pb collisions at 2.76 TeV}.
\newblock {\em Phys. Rev. Lett.}, 105:252302, 2010.

\bibitem{CMS:2012zex}
S.~Chatrchyan et~al.
\newblock {Measurement of the elliptic anisotropy of charged particles produced
  in PbPb collisions at $\sqrt{s}_{NN}$=2.76 TeV}.
\newblock {\em Phys. Rev. C}, 87(1):014902, 2013.

\bibitem{Ollitrault:1992bk}
Jean-Yves Ollitrault.
\newblock {Anisotropy as a signature of transverse collective flow}.
\newblock {\em Phys. Rev. D}, 46:229--245, 1992.

\bibitem{Rischke:1995ir}
Dirk~H. Rischke, S.~Bernard, and J.~A. Maruhn.
\newblock {Relativistic hydrodynamics for heavy ion collisions. 1. General
  aspects and expansion into vacuum}.
\newblock {\em Nucl. Phys. A}, 595:346--382, 1995.

\bibitem{Sorge:1996pc}
H.~Sorge.
\newblock {Elliptical flow: A Signature for early pressure in ultrarelativistic
  nucleus-nucleus collisions}.
\newblock {\em Phys. Rev. Lett.}, 78:2309--2312, 1997.

\bibitem{Bass:1998vz}
S.A. Bass, M.~Gyulassy, H.~Stoecker, and W.~Greiner.
\newblock {Signatures of quark gluon plasma formation in high-energy heavy ion
  collisions: A Critical review}.
\newblock {\em J. Phys. G}, 25:R1--R57, 1999.

\bibitem{Aguiar:2001ac}
C.~E. Aguiar, Y.~Hama, T.~Kodama, and T.~Osada.
\newblock {Event-by-event fluctuations in hydrodynamical description of heavy
  ion collisions}.
\newblock {\em Nucl. Phys. A}, 698:639--642, 2002.

\bibitem{Shuryak:2003xe}
E.~Shuryak.
\newblock {Why does the quark gluon plasma at RHIC behave as a nearly ideal
  fluid?}
\newblock {\em Prog. Part. Nucl. Phys.}, 53:273--303, 2004.

\bibitem{Gyulassy:2004zy}
M.~Gyulassy and L.~McLerran.
\newblock {New forms of QCD matter discovered at RHIC}.
\newblock {\em Nucl. Phys. A}, 750:30--63, 2005.

\bibitem{Broniowski:2007ft}
W.~Broniowski, P.~Bozek, and M.~Rybczynski.
\newblock {Fluctuating initial conditions in heavy-ion collisions from the
  Glauber approach}.
\newblock {\em Phys. Rev. C}, 76:054905, 2007.

\bibitem{Andrade:2008xh}
R.~P.~G. Andrade, F.~Grassi, Y.~Hama, T.~Kodama, and W.~L. Qian.
\newblock {Importance of Granular Structure in the Initial Conditions for the
  Elliptic Flow}.
\newblock {\em Phys. Rev. Lett.}, 101:112301, 2008.

\bibitem{Hirano:2009ah}
T.~Hirano and Y.~Nara.
\newblock {Eccentricity fluctuation effects on elliptic flow in relativistic
  heavy ion collisions}.
\newblock {\em Phys. Rev. C}, 79:064904, 2009.

\bibitem{Schenke:2010rr}
B.~Schenke, Sangyong Jeon, and C.~Gale.
\newblock {Elliptic and triangular flow in event-by-event (3+1)D viscous
  hydrodynamics}.
\newblock {\em Phys. Rev. Lett.}, 106:042301, 2011.

\bibitem{Qiu:2011iv}
Zhi Qiu and Ulrich~W. Heinz.
\newblock {Event-by-event shape and flow fluctuations of relativistic heavy-ion
  collision fireballs}.
\newblock {\em Phys. Rev. C}, 84:024911, 2011.

\bibitem{Heinz:2013th}
U.~Heinz and R.~Snellings.
\newblock {Collective flow and viscosity in relativistic heavy-ion collisions}.
\newblock {\em Ann. Rev. Nucl. Part. Sci.}, 63:123--151, 2013.

\bibitem{Huovinen:2013wma}
P.~Huovinen.
\newblock {Hydrodynamics at RHIC and LHC: What have we learned?}
\newblock {\em Int. J. Mod. Phys. E}, 22:1330029, 2013.

\bibitem{Gale:2013da}
C.~Gale, Sangyong Jeon, and B.~Schenke.
\newblock {Hydrodynamic Modeling of Heavy-Ion Collisions}.
\newblock {\em Int. J. Mod. Phys. A}, 28:1340011, 2013.

\bibitem{Bozek:2013uha}
P.~Bozek and W.~Broniowski.
\newblock {Collective dynamics in high-energy proton-nucleus collisions}.
\newblock {\em Phys. Rev. C}, 88(1):014903, 2013.

\bibitem{Qin:2013bha}
Guang-You Qin and B.~M\"uller.
\newblock {Elliptic and triangular flow anisotropy in deuteron-gold collisions
  at $\sqrt{s_{NN}}=200$ GeV at RHIC and in proton-lead collisions at
  $\sqrt{s_{NN}}=5.02$ TeV at the LHC}.
\newblock {\em Phys. Rev. C}, 89(4):044902, 2014.

\bibitem{Dusling:2015gta}
K.~Dusling, Wei Li, and B.~Schenke.
\newblock {Novel collective phenomena in high-energy proton\textendash{}proton
  and proton\textendash{}nucleus collisions}.
\newblock {\em Int. J. Mod. Phys. E}, 25(01):1630002, 2016.

\bibitem{Romatschke:2017ejr}
P.~Romatschke and U.~Romatschke.
\newblock {\em {Relativistic Fluid Dynamics In and Out of Equilibrium}}.
\newblock Cambridge Monographs on Mathematical Physics. Cambridge University
  Press, 5 2019.

\bibitem{Weller:2017tsr}
R.~D. Weller and P.~Romatschke.
\newblock {One fluid to rule them all: viscous hydrodynamic description of
  event-by-event central p+p, p+Pb and Pb+Pb collisions at $\sqrt{s}=5.02$
  TeV}.
\newblock {\em Phys. Lett. B}, 774:351--356, 2017.

\bibitem{Zhao:2020wcd}
Wenbin Zhao, Che~Ming Ko, Yu-Xin Liu, Guang-You Qin, and Huichao Song.
\newblock {Probing the Partonic Degrees of Freedom in High-Multiplicity $p-Pb$
  collisions at $\sqrt {s_{NN}}$ = 5.02 TeV}.
\newblock {\em Phys. Rev. Lett.}, 125(7):072301, 2020.

\bibitem{Song:2010mg}
Huichao Song, Steffen~A. Bass, Ulrich Heinz, Tetsufumi Hirano, and Chun Shen.
\newblock {200 A GeV Au+Au collisions serve a nearly perfect quark-gluon
  liquid}.
\newblock {\em Phys. Rev. Lett.}, 106:192301, 2011.
\newblock [Erratum: Phys.Rev.Lett. 109, 139904 (2012)].

\bibitem{Bernhard:2019bmu}
Jonah~E. Bernhard, J.~Scott Moreland, and Steffen~A. Bass.
\newblock {Bayesian estimation of the specific shear and bulk viscosity of
  quark\textendash{}gluon plasma}.
\newblock {\em Nature Phys.}, 15(11):1113--1117, 2019.

\bibitem{Gyulassy:1981nq}
M.~Gyulassy, K.~A. Frankel, and Horst Stoecker.
\newblock {DO NUCLEI FLOW AT HIGH-ENERGIES?}
\newblock {\em Phys. Lett. B}, 110:185--188, 1982.

\bibitem{Gustafsson:1984ka}
H.~A. Gustafsson et~al.
\newblock {Collective Flow Observed in Relativistic Nuclear Collisions}.
\newblock {\em Phys. Rev. Lett.}, 52:1590--1593, 1984.

\bibitem{Lisa:2000ip}
Michael~Annan Lisa, Ulrich~W. Heinz, and Urs~Achim Wiedemann.
\newblock {Tilted pion sources from azimuthally sensitive HBT interferometry}.
\newblock {\em Phys. Lett. B}, 489:287--292, 2000.

\bibitem{Voloshin:1994mz}
S.~Voloshin and Y.~Zhang.
\newblock {Flow study in relativistic nuclear collisions by Fourier expansion
  of Azimuthal particle distributions}.
\newblock {\em Z. Phys. C}, 70:665--672, 1996.

\bibitem{Bilandzic:2010jr}
A.~Bilandzic, R.~Snellings, and S.~Voloshin.
\newblock {Flow analysis with cumulants: Direct calculations}.
\newblock {\em Phys. Rev. C}, 83:044913, 2011.

\bibitem{STAR:2004jwm}
J.~Adams et~al.
\newblock {Azimuthal anisotropy in Au+Au collisions at s(NN)**(1/2) = 200-GeV}.
\newblock {\em Phys. Rev. C}, 72:014904, 2005.

\bibitem{STAR:2014clz}
L.~Adamczyk et~al.
\newblock {Beam-Energy Dependence of the Directed Flow of Protons, Antiprotons,
  and Pions in Au+Au Collisions}.
\newblock {\em Phys. Rev. Lett.}, 112(16):162301, 2014.

\bibitem{STAR:2017okv}
L.~Adamczyk et~al.
\newblock {Beam-Energy Dependence of Directed Flow of $\Lambda$,
  $\bar{\Lambda}$, $K^\pm$, $K^0_s$ and $\phi$ in Au+Au Collisions}.
\newblock {\em Phys. Rev. Lett.}, 120(6):062301, 2018.

\bibitem{STAR:2019clv}
J.~Adam et~al.
\newblock {First Observation of the Directed Flow of $D^{0}$ and
  $\overline{D^0}$ in Au+Au Collisions at $\sqrt{s_{\rm NN}}$ = 200 GeV}.
\newblock {\em Phys. Rev. Lett.}, 123(16):162301, 2019.

\bibitem{ALICE:2019sgg}
S.~Acharya et~al.
\newblock {Probing the effects of strong electromagnetic fields with
  charge-dependent directed flow in Pb-Pb collisions at the LHC}.
\newblock {\em Phys. Rev. Lett.}, 125(2):022301, 2020.

\bibitem{STAR:2019vcp}
Jaroslav Adam et~al.
\newblock {Bulk properties of the system formed in $Au+Au$ collisions at
  $\sqrt{s_{\mathrm{NN}}}$ =14.5 GeV at the BNL STAR detector}.
\newblock {\em Phys. Rev. C}, 101(2):024905, 2020.

\bibitem{Singha:2016mna}
S.~Singha, P.~Shanmuganathan, and D.~Keane.
\newblock {The first moment of azimuthal anisotropy in nuclear collisions from
  AGS to LHC energies}.
\newblock {\em Adv. High Energy Phys.}, 2016:2836989, 2016.

\bibitem{Nara:2016phs}
Y.~Nara, H.~Niemi, A.~Ohnishi, and H.~St\"ocker.
\newblock {Examination of directed flow as a signature of the softest point of
  the equation of state in QCD matter}.
\newblock {\em Phys. Rev. C}, 94(3):034906, 2016.

\bibitem{Chatterjee:2017ahy}
S.~Chatterjee and P.~Bo\.zek.
\newblock {Large directed flow of open charm mesons probes the three
  dimensional distribution of matter in heavy ion collisions}.
\newblock {\em Phys. Rev. Lett.}, 120(19):192301, 2018.

\bibitem{Zhang:2018wlk}
Chao Zhang, Jiamin Chen, Xiaofeng Luo, Feng Liu, and Y.~Nara.
\newblock {Beam energy dependence of the squeeze-out effect on the directed and
  elliptic flow in Au + Au collisions in the high baryon density region}.
\newblock {\em Phys. Rev. C}, 97(6):064913, 2018.

\bibitem{Guo:2017mkf}
Chong-Qiang Guo, Chun-Jian Zhang, and Jun Xu.
\newblock {Revisiting directed flow in relativistic heavy-ion collisions from a
  multiphase transport model}.
\newblock {\em Eur. Phys. J. A}, 53(12):233, 2017.

\bibitem{Parida:2022lmt}
Tribhuban Parida and Sandeep Chatterjee.
\newblock {Splitting of elliptic flow in a tilted fireball}.
\newblock {\em arXiv: 2204.02345}.

\bibitem{Adil:2005qn}
A.~Adil and M.~Gyulassy.
\newblock {3D jet tomography of twisted strongly coupled quark gluon plasmas}.
\newblock {\em Phys. Rev. C}, 72:034907, 2005.

\bibitem{Bozek:2010bi}
P.~Bozek and I.~Wyskiel.
\newblock {Directed flow in ultrarelativistic heavy-ion collisions}.
\newblock {\em Phys. Rev. C}, 81:054902, 2010.

\bibitem{Chen:2019qzx}
Baoyi Chen, Maoxin Hu, Huanyu Zhang, and Jiaxing Zhao.
\newblock {Probe the tilted Quark-Gluon Plasma with charmonium directed flow}.
\newblock {\em Phys. Lett. B}, 802:135271, 2020.

\bibitem{Shen:2020jwv}
Chun Shen and S.~Alzhrani.
\newblock {Collision-geometry-based 3D initial condition for relativistic
  heavy-ion collisions}.
\newblock {\em Phys. Rev. C}, 102(1):014909, 2020.

\bibitem{Ryu:2021lnx}
Sangwook Ryu, Vahidin Jupic, and Chun Shen.
\newblock {Probing early-time longitudinal dynamics with the
  \ensuremath{\Lambda} hyperon's spin polarization in relativistic heavy-ion
  collisions}.
\newblock {\em Phys. Rev. C}, 104(5):054908, 2021.

\bibitem{Chatterjee:2018lsx}
S.~Chatterjee and P.~Bozek.
\newblock {Interplay of drag by hot matter and electromagnetic force on the
  directed flow of heavy quarks}.
\newblock {\em Phys. Lett. B}, 798:134955, 2019.

\bibitem{Beraudo:2021ont}
A.~Beraudo, A.~De~Pace, M.~Monteno, M.~Nardi, and F.~Prino.
\newblock {Rapidity dependence of heavy-flavour production in heavy-ion
  collisions within a full 3+1 transport approach: quenching, elliptic and
  directed flow}.
\newblock {\em JHEP}, 05:279, 2021.

\bibitem{Bozek:2022svy}
Piotr Bozek.
\newblock {Splitting of proton-antiproton directed flow in relativistic
  heavy-ion collisions}.
\newblock {\em arXiv: ~2207.04927}.

\bibitem{STAR:2021mii}
Mohamed Abdallah et~al.
\newblock {Search for the chiral magnetic effect with isobar collisions at
  $\sqrt {s_{NN}}$=200 GeV by the STAR Collaboration at the BNL Relativistic
  Heavy Ion Collider}.
\newblock {\em Phys. Rev. C}, 105(1):014901, 2022.

\bibitem{Fukushima:2008xe}
Kenji Fukushima, Dmitri~E. Kharzeev, and Harmen~J. Warringa.
\newblock {The Chiral Magnetic Effect}.
\newblock {\em Phys. Rev. D}, 78:074033, 2008.

\bibitem{Skokov:2009qp}
V.~Skokov, A.~Yu. Illarionov, and V.~Toneev.
\newblock {Estimate of the magnetic field strength in heavy-ion collisions}.
\newblock {\em Int. J. Mod. Phys. A}, 24:5925--5932, 2009.

\bibitem{Deng:2016knn}
Wei-Tian Deng, Xu-Guang Huang, Guo-Liang Ma, and Gang Wang.
\newblock {Test the chiral magnetic effect with isobaric collisions}.
\newblock {\em Phys. Rev. C}, 94:041901, 2016.

\bibitem{Zhao:2019crj}
Xin-Li Zhao, Guo-Liang Ma, and Yu-Gang Ma.
\newblock {Impact of magnetic-field fluctuations on measurements of the chiral
  magnetic effect in collisions of isobaric nuclei}.
\newblock {\em Phys. Rev. C}, 99(3):034903, 2019.

\bibitem{Nijs:2021kvn}
Govert Nijs and Wilke van~der Schee.
\newblock {Inferring nuclear structure from heavy isobar collisions using
  Trajectum}.
\newblock {\em arXiv: 2112.13771}.

\bibitem{Zhang:2022fou}
Chunjian Zhang, Somadutta Bhatta, and Jiangyong Jia.
\newblock {Ratios of collective flow observables in high-energy isobar
  collisions are insensitive to final state interactions}.
\newblock {\em arXiv: 2206.01943}.

\bibitem{Zhang:2021kxj}
Chunjian Zhang and Jiangyong Jia.
\newblock {Evidence of Quadrupole and Octupole Deformations in Zr96+Zr96 and
  Ru96+Ru96 Collisions at Ultrarelativistic Energies}.
\newblock {\em Phys. Rev. Lett.}, 128(2):022301, 2022.

\bibitem{Jia:2021oyt}
Jiangyong Jia and Chun-Jian Zhang.
\newblock {Scaling approach to nuclear structure in high-energy heavy-ion
  collisions}.
\newblock {\em arXiv: 2111.15559}.

\bibitem{Xu:2021uar}
Hao-jie Xu, Wenbin Zhao, Hanlin Li, Ying Zhou, Lie-Wen Chen, and Fuqiang Wang.
\newblock {Probing nuclear structure with mean transverse momentum in
  relativistic isobar collisions}.
\newblock {\em arXiv: 2111.14812}.

\bibitem{Li:2022bhl}
Fei Li, Yu-Gang Ma, Song Zhang, Guo-Liang Ma, and Qi-Ye Shou.
\newblock {Impact of nuclear structure on the CME background in $^{96}_{44}$Ru
  + $^{96}_{44}$Ru and $^{96}_{40}$Zr + $^{96}_{40}$Zr collisions at
  $\sqrt{s_{NN}}$ = 7.7 $\sim$ 200 GeV from a multiphase transport model}.
\newblock {\em arXiv: 2201.10994}.

\bibitem{Zhao:2022grq}
Xin-Li Zhao and Guo-Liang Ma.
\newblock {Search for the chiral magnetic effect in collisions between two
  isobars with deformed and neutron-rich nuclear structures}.
\newblock {\em Phys. Rev. C}, 106(3):034909, 2022.

\bibitem{Ma:2022dbh}
Yu-Gang Ma and Song Zhang.
\newblock {Influence of nuclear structure in relativistic heavy-ion
  collisions}.
\newblock {\em arXiv: 2206.08218}.

\bibitem{Jia:2022qrq}
Jiangyong Jia, Giuliano Giacalone, and Chunjian Zhang.
\newblock {Precision tests of the nonlinear mode coupling of anisotropic flow
  via high-energy collisions of isobars}.
\newblock {\em arXiv: 2206.07184}.

\bibitem{Pang:2016igs}
Long-Gang Pang, H.~Petersen, Qun Wang, and Xin-Nian Wang.
\newblock {Vortical Fluid and $\Lambda$ Spin Correlations in High-Energy
  Heavy-Ion Collisions}.
\newblock {\em Phys. Rev. Lett.}, 117(19):192301, 2016.

\bibitem{Pang:2018zzo}
Long-Gang Pang, H.~Petersen, and Xin-Nian Wang.
\newblock {Pseudorapidity distribution and decorrelation of anisotropic flow
  within the open-computing-language implementation CLVisc hydrodynamics}.
\newblock {\em Phys. Rev. C}, 97(6):064918, 2018.

\bibitem{Wu:2018cpc}
Xiang-Yu Wu, Long-Gang Pang, Guang-You Qin, and Xin-Nian Wang.
\newblock {Longitudinal fluctuations and decorrelations of anisotropic flows at
  energies available at the CERN Large Hadron Collider and at the BNL
  Relativistic Heavy Ion Collider}.
\newblock {\em Phys. Rev. C}, 98(2):024913, 2018.

\bibitem{Xu:2017zcn}
Hao-Jie Xu, Xiaobao Wang, Hanlin Li, Jie Zhao, Zi-Wei Lin, Caiwan Shen, and
  Fuqiang Wang.
\newblock {Importance of isobar density distributions on the chiral magnetic
  effect search}.
\newblock {\em Phys. Rev. Lett.}, 121(2):022301, 2018.

\bibitem{Li:2018oec}
Hanlin Li, Hao-jie Xu, Jie Zhao, Zi-Wei Lin, Hanzhong Zhang, Xiaobao Wang,
  Caiwan Shen, and Fuqiang Wang.
\newblock {Multiphase transport model predictions of isobaric collisions with
  nuclear structure from density functional theory}.
\newblock {\em Phys. Rev. C}, 98(5):054907, 2018.

\bibitem{Xu:2021vpn}
Hao-jie Xu, Hanlin Li, Xiaobao Wang, Caiwan Shen, and Fuqiang Wang.
\newblock {Determine the neutron skin type by relativistic isobaric
  collisions}.
\newblock {\em Phys. Lett. B}, 819:136453, 2021.

\bibitem{Abelev:2008jga}
B.I. Abelev et~al.
\newblock {System-size independence of directed flow at the Relativistic
  Heavy-Ion Collider}.
\newblock {\em Phys. Rev. Lett.}, 101:252301, 2008.

\bibitem{qm2022ruzr}
A.~I.~Sheikh for~the STAR~Collaboration.
\newblock {Splitting of directed flow for identified light hadrons ($\pi$, K,
  p) and strange baryons ($\Xi$, $\Omega$) in Au+Au and isobar collisions at
  STAR}.
\newblock {\em 29th Conference on Ultra-telativistic Nucleus-Nucleus Collisions
  -- Quark Matter 2022}.

\bibitem{Jiang:2021foj}
Ze-Fang Jiang, C.~B. Yang, and Qi~Peng.
\newblock {Directed flow of charged particles within idealized viscous
  hydrodynamics at energies available at the BNL Relativistic Heavy Ion
  Collider and at the CERN Large Hadron Collider}.
\newblock {\em Phys. Rev. C}, 104(6):064903, 2021.

\bibitem{Jiang:2021ajc}
Ze-Fang Jiang, Shanshan Cao, Xiang-Yu Wu, C.~B. Yang, and Ben-Wei Zhang.
\newblock {Longitudinal distribution of initial energy density and directed
  flow of charged particles in relativistic heavy-ion collisions}.
\newblock {\em Phys. Rev. C}, 105(3):034901, 2022.

\bibitem{Jiang:2022uoe}
Ze-Fang Jiang, Shanshan Cao, Wen-Jing Xing, Xiang-Yu Wu, C.~B. Yang, and
  Ben-Wei Zhang.
\newblock {Probing the initial longitudinal density profile and electromagnetic
  field in ultrarelativistic heavy-ion collisions with heavy quarks}.
\newblock {\em Phys. Rev. C}, 105(5):054907, 2022.

\bibitem{Li:2022pyw}
Xiaowen Li, Ze-Fang Jiang, Shanshan Cao, and Jian Deng.
\newblock {Evolution of global polarization in relativistic heavy-ion
  collisions within a perturbative approach}.
\newblock {\em arXiv: 2205.02409}.

\bibitem{Deng:2018dut}
Wei-Tian Deng, Xu-Guang Huang, Guo-Liang Ma, and Gang Wang.
\newblock {Predictions for isobaric collisions at $\sqrt{s_{_{\rm NN}}}$ = 200
  GeV from a multiphase transport model}.
\newblock {\em Phys. Rev. C}, 97(4):044901, 2018.

\bibitem{Loizides:2017ack}
C.~Loizides, J.~Kamin, and D.~d'Enterria.
\newblock {Improved Monte Carlo Glauber predictions at present and future
  nuclear colliders}.
\newblock {\em Phys. Rev. C}, 97(5):054910, 2018.
\newblock [Erratum: Phys.Rev.C 99, 019901 (2019)].

\bibitem{Bozek:2011ua}
Piotr Bozek.
\newblock {Flow and interferometry in 3+1 dimensional viscous hydrodynamics}.
\newblock {\em Phys. Rev. C}, 85:034901, 2012.

\bibitem{Chen:2017zte}
Wei Chen, Shanshan Cao, Tan Luo, Long-Gang Pang, and Xin-Nian Wang.
\newblock {Effects of jet-induced medium excitation in $\gamma$-hadron
  correlation in A+A collisions}.
\newblock {\em Phys. Lett. B}, 777:86--90, 2018.

\bibitem{He:2018gks}
Yayun He, Long-Gang Pang, and Xin-Nian Wang.
\newblock {Bayesian extraction of jet energy loss distributions in heavy-ion
  collisions}.
\newblock {\em Phys. Rev. Lett.}, 122(25):252302, 2019.

\bibitem{Jiang:2020big}
Ze~Fang Jiang, Duan She, C.B. Yang, and Defu Hou.
\newblock {Perturbation solutions of relativistic viscous hydrodynamics
  forlongitudinally expanding fireballs}.
\newblock {\em Chin. Phys. C}, 44(8):084107, 2020.

\bibitem{Jiang:2018qxd}
Ze~Fang Jiang, C.B. Yang, Chi Ding, and Xiang-Yu Wu.
\newblock {Pseudo-rapidity distribution from a perturbative solution of viscous
  hydrodynamics for heavy ion collisions at RHIC and LHC}.
\newblock {\em Chin. Phys. C}, 42(12):123103, 2018.

\bibitem{Denicol:2012cn}
G.S. Denicol, H.~Niemi, E.~Molnar, and D.H. Rischke.
\newblock {Derivation of transient relativistic fluid dynamics from the
  Boltzmann equation}.
\newblock {\em Phys. Rev. D}, 85:114047, 2012.
\newblock [Erratum: Phys.Rev.D 91, 039902 (2015)].

\bibitem{Romatschke:2009im}
P.~Romatschke.
\newblock {New Developments in Relativistic Viscous Hydrodynamics}.
\newblock {\em Int. J. Mod. Phys. E}, 19:1--53, 2010.

\bibitem{Borsanyi:2013bia}
S.~Borsanyi, Z.~Fodor, C.~Hoelbling, S.~D. Katz, S.~Krieg, and K.~K. Szabo.
\newblock {Full result for the QCD equation of state with 2+1 flavors}.
\newblock {\em Phys. Lett. B}, 730:99--104, 2014.

\bibitem{Zhao:2021vmu}
Wenbin Zhao, Weiyao Ke, Wei Chen, Tan Luo, and Xin-Nian Wang.
\newblock {From Hydrodynamics to Jet Quenching, Coalescence, and Hadron
  Cascade: A Coupled Approach to Solving the RAA\ensuremath{\otimes}v2 Puzzle}.
\newblock {\em Phys. Rev. Lett.}, 128(2):022302, 2022.

\bibitem{Wu:2021fjf}
Xiang-Yu Wu, Guang-You Qin, Long-Gang Pang, and Xin-Nian Wang.
\newblock {(3+1)-D viscous hydrodynamics at finite net baryon density:
  Identified particle spectra, anisotropic flows, and flow fluctuations across
  energies relevant to the beam-energy scan at RHIC}.
\newblock {\em Phys. Rev. C}, 105(3):034909, 2022.

\bibitem{Cooper:1974mv}
F.~Cooper and G.~Frye.
\newblock {Comment on the Single Particle Distribution in the Hydrodynamic and
  Statistical Thermodynamic Models of Multiparticle Production}.
\newblock {\em Phys. Rev. D}, 10:186, 1974.

\bibitem{Alver:2010ck}
B.~Alver et~al.
\newblock {Phobos results on charged particle multiplicity and pseudorapidity
  distributions in Au+Au, Cu+Cu, d+Au, and p+p collisions at ultra-relativistic
  energies}.
\newblock {\em Phys. Rev. C}, 83:024913, 2011.

\bibitem{Raman:2001nnq}
S.~Raman, C.~W.~G. Nestor, Jr, and P.~Tikkanen.
\newblock {Transition probability from the ground to the first-excited 2+ state
  of even-even nuclides}.
\newblock {\em Atom. Data Nucl. Data Tabl.}, 78:1--128, 2001.

\bibitem{Pritychenko:2013gwa}
B.~Pritychenko, M.~Birch, B.~Singh, and M.~Horoi.
\newblock {Tables of E2 Transition Probabilities from the first $2^{+}$ States
  in Even-Even Nuclei}.
\newblock {\em Atom. Data Nucl. Data Tabl.}, 107:1--139, 2016.
\newblock [Erratum: Atom.Data Nucl.Data Tabl. 114, 371--374 (2017)].

\bibitem{Moller:1993ed}
P.~Moller, J.~R. Nix, W.~D. Myers, and W.~J. Swiatecki.
\newblock {Nuclear ground state masses and deformations}.
\newblock {\em Atom. Data Nucl. Data Tabl.}, 59:185--381, 1995.

\bibitem{Jia:2021tzt}
Jiangyong Jia.
\newblock {Shape of atomic nuclei in heavy ion collisions}.
\newblock {\em Phys. Rev. C}, 105(1):014905, 2022.

\bibitem{Teaney:2010vd}
Derek Teaney and Li~Yan.
\newblock {Triangularity and Dipole Asymmetry in Heavy Ion Collisions}.
\newblock {\em Phys. Rev. C}, 83:064904, 2011.

\bibitem{Luzum:2010fb}
Matthew Luzum and Jean-Yves Ollitrault.
\newblock {Directed flow at midrapidity in heavy-ion collisions}.
\newblock {\em Phys. Rev. Lett.}, 106:102301, 2011.

\bibitem{Gale:2012rq}
Charles Gale, Sangyong Jeon, Bj\"orn Schenke, Prithwish Tribedy, and Raju
  Venugopalan.
\newblock {Event-by-event anisotropic flow in heavy-ion collisions from
  combined Yang-Mills and viscous fluid dynamics}.
\newblock {\em Phys. Rev. Lett.}, 110(1):012302, 2013.

\bibitem{Inghirami:2019mkc}
G.~Inghirami, M.~Mace, Y.~Hirono, L.~Del~Zanna, D.~E. Kharzeev, and
  M.~Bleicher.
\newblock {Magnetic fields in heavy ion collisions: flow and charge transport}.
\newblock {\em Eur. Phys. J. C}, 80(3):293, 2020.

\bibitem{Gursoy:2014aka}
U.~Gursoy, D.~E. Kharzeev, and K.~Rajagopal.
\newblock {Magnetohydrodynamics, charged currents and directed flow in heavy
  ion collisions}.
\newblock {\em Phys. Rev. C}, 89(5):054905, 2014.

\bibitem{Gursoy:2018yai}
U.~G\"ursoy, D.~E. Kharzeev, E.~Marcus, K.~Rajagopal, and Chun Shen.
\newblock {Charge-dependent Flow Induced by Magnetic and Electric Fields in
  Heavy Ion Collisions}.
\newblock {\em Phys. Rev. C}, 98(5):055201, 2018.

\end{thebibliography}

\end{document}